\documentclass[11pt]{article}
\usepackage{amsmath, mathrsfs, amsfonts, amssymb, amsthm, newlfont, graphicx, yfonts}

\begin{document}

\title{{\bf The Extended Estabrook-Wahlquist Method}}

\author{Matthew Russo and S. Roy Choudhury\\  \small  Department of Mathematics, University of Central Florida, Orlando, FL  32816-1364 USA\\
Corresponding author email: choudhur@cs.ucf.edu }      
\date{\today}       
\maketitle

\centerline{ \noindent\Large\textbf{Abstract}}
Variable Coefficient Korteweg de Vries (vcKdV), Modified Korteweg de Vries (vcMKdV), and nonlinear Schrodinger (NLS) equations have a long history dating from their derivation in various applications. A technique based on extended Lax Pairs has been devised recently to derive variable-coefficient generalizations of various Lax-integrable NLPDE hierarchies. The resulting Lax- or S-integrable NLPDEs have both time- AND space-dependent
coefficients, and are thus more general than almost all cases considered earlier via other methods such as the
Painlev\'e Test, Bell Polynomials, and various similarity methods.

However, this technique, although operationally effective, has the significant disadvantage that, for any integrable system with spatiotemporally varying coefficients, one must 'guess' a generalization of the structure of the known Lax Pair for the corresponding system with constant coefficients. Motivated by the somewhat arbitrary nature of the above procedure, we embark in this paper on an attempt to systematize the derivation of Lax-integrable sytems with variable coefficients. An ideal approach would be a method which does not require knowledge of the Lax pair to an associated constant coefficient system, and also involves little to no guesswork. Hence we attempt to apply the Estabrook-Wahlquist (EW) prolongation technique, a relatively self-consistent procedure requiring little prior infomation. However, this immediately requires that the technique be significantly generalized or broadened in several different ways, including
solving matrix partial differential equations instead of algebraic ones as the structure of the Lax Pair is deduced systematically following the standard Lie-algebraic procedure of proceeding downwards from the coefficient of the highest derivative. The same is true while finding the explicit forms for the various 'coefficient' matrices which occur in the procedure, and which must satisfy the various constraint equations which result at various stages of the calculation. 

The new and extended EW technique whch results is illustrated by algorithmically deriving generalized Lax-integrable versions of the NLS, generalized fifth-order KdV, MKdV, and derivative nonlinear Schrodinger (DNLS) equations.\\
\\

Key Words: extended Estabrook-Wahlquist Method, Generalizing Lax or S-integrable equations, spatially and temporally-dependent coefficients.

\section{Introduction}

Variable Coefficient Korteweg de Vries (vcKdV) and Modified Korteweg de Vries (vcMKdV)equations have a long history dating from their derivation in various applications\cite{K1}-\cite{K10}. However, almost all studies, including those which derived exact solutions by a variety of techniques,
as well as those which considered integrable sub-cases and various integrability properties by methods such as Painlev\'e analysis, Hirota's method, and Bell Polynomials treat vcKdV equations with coefficients which are functions of the time only. For instance, for generalized variable coefficient NLS (vcNLS) equations, a particular coefficient is usually taken to be a function of $x$ \cite{K11}, as has also been sometimes done for vcMKdV equations\cite{K12}. 

The papers \cite{K13}-\cite{Khawaja}, and some of the references therein
, are somewhat of an exception in that they treat vcNLS equations having coefficients with general $x$ and $t$ dependences. Variational principles, solutions, and other integrability properties have also been considered for some of the above variable coefficient NLPDEs in cases with time-dependent coefficients. The technique in \cite{Khawaja} has recently been extended\cite{Lecce} to build integrable families of KdV and MKdV equations with
both spatially and temporally varying coefficents.

However, this technique, although operationally effective, has the significant disadvantage that, for any integrable system with spatiotemporally varying coefficients, one must 'guess' a generalization of the structure of the known Lax Pair for the corresponding system with constant coefficients. This involves
replacing constants in the Lax Pair for the constant coefficient integrable system, including powers of the spectral parameter, by functions. Provided that one has guessed correctly and generalized the constant coefficient system's Lax Pair sufficiently, and this is  of course hard to be sure of 'a priori', one may then
proceed to systematically deduce the Lax Pair for the corresponding variable-coefficient integrable system \cite{Lecce}.

Motivated by the somewhat arbitrary nature of the above procedure, we embark in this paper on an attempt to systematize the derivation of Lax-integrable sytems with variable coefficients. Of the many techniques which have been employed for constant coefficient integrable systems, the Estabrook-Wahlquist (EW) prolongation technique \cite{EW1}-\cite{EW4}is among the most self-contained. The method directly proceeds to attempt construction of the Lax Pair or linear spectral problem, whose compatibility condition is the integrable system under discussion. While not at all guaranteed to work, any successful implementation of the technique means that Lax-integrability has already been verified during the procedure, and in addition the Lax Pair is algorithmically obtained. If the technique fails, that does not necessarily imply non-integrability  of the equation contained in the compatibility condition of the assumed Lax Pair. It may merely mean that some of the starting assumptions may not be appropriate or general enough.

In applications, the coefficients of vcKdV equations may include spatial dependence, in addition to the temporal variations that have been extensively considered using a variety of techniques. Both for this reason, as well as
for their general mathematical interest, extending integrable hierarchies of nonlinear PDEs (NLPDEs) to include {\it both} spatial and temporal dependence of the coefficients is worthwhile. Hence we attempt to apply the Estabrook-Wahlquist (EW) technique to generate a variety of such integrable systems with such spatiotemporally varying coefficients. However, this immediately requires that the technique be significantly generalized or broadened in several different ways which we outline
in the following section, before illustrating this new and extended method with a variety of examples.

The remainder of this paper is organized as follows. In section 2, we lay out the extensions required to apply the EW procedure to Lax-integrbale systems
with spatiotemporally varying coefficients. Sections 3-6 the illustrate the method in detail for Lax-integrable versions of the nonlinear Schroedinger (NLS),
generalized Korteweg-deVries (KdV), modified Korteweg-deVries (KdV), and derivative nonlinear Schroedinger (DNSL) equations respectively, each with spatiotemporally varying coefficients. In Sections 3-5, we also illustrate that this generalized EEW procedure algorithmically generates the same results
as those obtained in a more ad hoc manner in \cite{Khawaja}-\cite{Lecce} by generalizing Lax Pairs for the corresponding constant coefficient integrable systems by guesswork.

\section{The extensions of the EW technique}

In the standard Estabrook-Wahlquist method one begins with a constant coefficient NLPDE and assumes an implicit dependence on $u(x,t)$ and its partial derivatives of the spatial and time evolution matrices ($\mathbb{F},\mathbb{G}$) involved in the linear scattering problem 
\[ \psi_{x} = \mathbb{F}\psi, \ \ \ \psi_{t} = \mathbb{G}\psi \]
The evolution matrices $\mathbb{F}$ and $\mathbb{G}$ are connected via a zero-curvature condition (independence of path in spatial and time evolution) derived by mandating $\psi_{xt} = \psi_{tx}$. That is, it requires
\[ \mathbb{F}_{t} - \mathbb{G}_{x} + [\mathbb{F},\mathbb{G}] = 0 \]
provided $u(x,t)$ satisfies the NLPDE. 

Considering the forms $\mathbb{F} = \mathbb{F}(u,u_{x},u_{t},\ldots,u_{mx,nt})$ and $\mathbb{G} = \mathbb{G}(u,u_{x},u_{t},\ldots,u_{kx,jt})$ for the space and time evolution matrices where $u_{px,qt} = \frac{\partial^{p+q}u}{\partial x^{p}\partial t^{q}}$ we see that this condition is equivalent to

\[ \sum_{m,n}{\mathbb{F}_{u_{mx,nt}}u_{mx,(n+1)t}} - \sum_{j,k}{\mathbb{G}_{u_{jx,kt}}u_{(j+1)x,kt}} + [\mathbb{F},\mathbb{G}] = 0 \]

\noindent
From here there is often a systematic approach\cite{EW1}-\cite{EW4} to determining the form for $\mathbb{F}$ and $\mathbb{G}$ which is outlined in \cite{EW3} and will be utilized in the examples to follow. 

Typically a valid choice for dependence on $u(x,t)$ and its partial derivatives is to take $\mathbb{F}$ to depend on all terms in the NLPDE for which there is a partial time derivative present. Similarly we may take $\mathbb{G}$ to depend on all terms for which there is a partial space derivative present. For example, given the Camassa-Holm equation,
\[ u_{t} + 2ku_{x} - u_{xxt} + 3uu_{x} - 2u_{x}u_{xx} - uu_{xxx} = 0, \]
one would consider $\mathbb{F} = \mathbb{F}(u,u_{xx})$ and $\mathbb{G} = \mathbb{G}(u,u_{x},u_{xx})$. Imposing compatibility allows one to determine the explicit form of $\mathbb{F}$ and $\mathbb{G}$ in a very algorithmic way. Additionally the compatibility condition induces a set of constraints on the coefficient matrices in $\mathbb{F}$ and $\mathbb{G}$. These coefficient matrices subject to the constraints generate a finite dimensional matrix Lie algebra.

In the extended Estabrook-Wahlquist method we allow for $\mathbb{F}$ and $\mathbb{G}$ to be functions of $t$, $x$, $u$ and the partial derivatives of $u$. Although the details change, the general procedure will remain essentially the same. We will begin by equating the coefficient of the highest partial derivative of the unknown function(s) to zero and work our way down until we have eliminated all partial derivatives of the unknown function(s). 

{\it This typically results in a large partial differential equation (in the standard Estabrook-Wahlquist method, this is an algebraic equation) which can be solved by equating the coefficients of the different powers of the unknown function(s) to zero.} This final step induces a set of constraints on the coefficient matrices in $\mathbb{F}$ and $\mathbb{G}$. {\it Another big difference which we will see in the examples comes in the final and, arguably, the hardest step. In the standard Estabrook-Wahlquist method the final step involves finding explicit forms for the set of coefficient matrices such that they satisfy the contraints derived in the procedure. Note these constraints are nothing more than a system of algebraic matrix equations. In the extended Estabrook-Wahlquist method these constraints will be in the form of matrix partial differential equations which can be used to derive an integrability condition on the coefficients in the NLPDE.}

As we are now letting $\mathbb{F}$ and $\mathbb{G}$ have explicit dependence on $x$ and $t$ and for notational clarity, it will be more convenient to consider the following version of the zero-curvature condition

\begin{equation}\label{ZCC}
\mbox{D}_{t}\mathbb{F} - \mbox{D}_{x}\mathbb{G} + [\mathbb{F},\mathbb{G}] = 0
\end{equation}

\noindent
where $\mbox{D}_{t}$ and $\mbox{D}_{x}$ are the total derivative operators on time and space, respectively. Recall the definition of the total derivative

\[ \mbox{D}_{y}f(y,z,u_{1}(y,z),u_{2}(y,z),\ldots,u_{n}(y,z)) = \frac{\partial f}{\partial y} + \frac{\partial f}{\partial u_{1}}\frac{\partial u_{1}}{\partial y} + \frac{\partial f}{\partial u
_{2}}\frac{\partial u_{2}}{\partial y} + \cdots + \frac{\partial f}{\partial u_{n}}\frac{\partial u_{n}}{\partial y} \]

\noindent
Thus we can write the compatibility condition as

\[ \mathbb{F}_{t} + \sum_{m,n}{\mathbb{F}_{u_{mx,nt}}u_{mx,(n+1)t}} - \mathbb{G}_{x} - \sum_{j,k}{\mathbb{G}_{u_{jx,kt}}u_{(j+1)x,kt}} + [\mathbb{F},\mathbb{G}] = 0 \]

\noindent
It is important to note that the subscripted $x$ and $t$ denotes the partial derivative with respect to only the $x$ and $t$ elements, respectively. That is, although $u$ and it's derivatives depend on $x$ and $t$ this will not invoke use of the chain rule as they are treated as independent variables. This will become more clear in the examples of the next section. 

Note that compatibility of the time and space evolution matrices will yield a set of constraints which contain the constant coefficient constraints as a subset. In fact, taking the variable coefficients to be the appropriate constants will yield exactly the Estabrook-Wahlquist results for the constant coefficient version of the NLPDE. That is, the constraints given by the Estabrook-Wahlquist method for a constant coefficient NLPDE are always a proper subset of the constraints given by a variable-coefficient version of the NLPDE. This can easily be shown. Letting $\mathbb{F}$ and $\mathbb{G}$ not depend explicitly on $x$ and $t$ and taking the coefficients in the NLPDE to be constant the zero-curvature condition as it is written above becomes

\[ \sum_{m,n}{\mathbb{F}_{u_{mx,nt}}u_{mx,(n+1)t}} - \sum_{j,k}{\mathbb{G}_{u_{jx,kt}}u_{(j+1)x,kt}} + [\mathbb{F},\mathbb{G}] = 0 \]

\noindent
which is exactly the standard Estabrook-Wahlquist method. 

The conditions derived via mandating $\eqref{ZCC}$ be satisfied upon solutions of the vc-NLPDE may be used to determine conditions on the coefficient matrices and variable-coefficients (present in the NLPDE). Successful closure of these conditions is equivalent to the system being S-integrable. A major advantage to using the Estabrook-Wahlquist method that carries forward with the extension is the fact that it requires little guesswork and yields quite general results. 

In Khawaja's method\cite{Khawaja}-\cite{Lecce} an educated guess is made for the structure of the variable-coefficient pde Lax pair based on the associated constant coefficient Lax pair. That is, Khawaja considered the matrices
\[ \mathbb{F} = U = \begin{bmatrix}
f_{1} + f_{2}q & f_{3} + f_{4}q \\
f_{5} + f_{6}r & f_{7} + f_{8}r
\end{bmatrix} \]
and
\[ \mathbb{G} = V = \begin{bmatrix}
g_{1} + g_{2}q + g_{3}q_{x} + g_{4}rq & g_{5} + g_{6}q + g_{7}q_{x} + g_{8}rq \\
g_{9} + g_{10}r + g_{11}r_{x} + g_{12}rq & g_{13} + g_{14}r + g_{15}r_{x} + g_{16}rq
\end{bmatrix} \]
where $f_{i}$ and $g_{i}$ unknown functions of $x$ and $t$ which satisfy conditions derived by mandating the zero-curvature condition be satisfied on solutions of the variable-coefficient NLPDE. In fact, in a previous paper Khawaja derives the associated Lax pair via a similar means where he begins with an even weaker assumption on the structure of the Lax pair. This Lax pair is omitted from the paper as it becomes clear the zero-curvature condition mandates many of the coefficients be zero. 

{\it An ideal approach would be a method which does not require knowledge of the Lax pair to an associated constant coefficient system and involves little to no guesswork. The extended Estabrook-Wahlquist does exactly this. It will be shown that the results obtained from Khawaja's method are in fact a special case of the extended Estabrook-Wahlquist method. }

We now proceed with the system Khawaja considered in his paper as our first example of the extended Estabrook Wahlquist method. 

\section{The Nonlinear Schrodinger (NLS) Equation}

The system is given by
\begin{subequations} \label{Khawaja}
\begin{align}
        fq_{xx} + gq^{2}r + (\upsilon+i\gamma)q + iq_{t}&=0,  \\
        fr_{xx} + gr^{2}q + (\upsilon-i\gamma)r - ir_{t}&=0,
\end{align}
\end{subequations}

\noindent
where $f,g,\upsilon,$ and $\gamma$ are functions of $x$ and $t$. 

Following with the procedure outlined above we choose

\[ \mathbb{F} = \mathbb{F}(x,t,q,r), \ \ \ \mathbb{G} = \mathbb{G}(x,t,r,q,r_{x},q_{x}) \]

\noindent
Compatibility requires these matrices satisfy the zero-curvature conditions given by $\eqref{ZCC}$. Plugging $\mathbb{F}$ and $\mathbb{G}$ into $\eqref{ZCC}$ we have

\begin{equation} \label{Step1}
\mathbb{F}_{t} + \mathbb{F}_{r}r_{t} + \mathbb{F}_{q}q_{t} - \mathbb{G}_{x} - \mathbb{G}_{r}r_{x} - \mathbb{G}_{q}q_{x} - \mathbb{G}_{r_{x}}r_{xx} - \mathbb{G}_{q_{x}}q_{xx} + \left[\mathbb{F},\mathbb{G}\right] = 0
\end{equation}

\noindent
Now requiring this be satisfied upon solutions of $\eqref{Khawaja}$ we follow the standard technique of eliminating $r_{t}$ and $u_{t}$ via $\eqref{Khawaja}$ to obtain

\begin{eqnarray} \nonumber
&& \mathbb{F}_{t} - i\mathbb{F}_{r}(fr_{xx} + gr^{2}q + (\upsilon-i\gamma)r) + i\mathbb{F}_{q}(fq_{xx} + gq^{2}r + (\upsilon+i\gamma)q) \\ \label{Step2}
&& - \mathbb{G}_{x} - \mathbb{G}_{r}r_{x} - \mathbb{G}_{q}q_{x} - \mathbb{G}_{r_{x}}r_{xx} - \mathbb{G}_{q_{x}}q_{xx} + \left[\mathbb{F},\mathbb{G}\right] = 0
\end{eqnarray}

Since $\mathbb{F}$ and $\mathbb{G}$ do not depend on $q_{xx}$ or $r_{xx}$ we collect the coefficients of $q_{xx}$ and $r_{xx}$ and equate them to zero. This requires

\begin{equation}
-if\mathbb{F}_{r} - \mathbb{G}_{r_{x}} = 0, \ \ \ \ \ if\mathbb{F}_{q} - \mathbb{G}_{q_{x}} = 0
\end{equation}

\noindent
Solving this linear system yields

\begin{equation}
\mathbb{G} = if(\mathbb{F}_{q}q_{x} - \mathbb{F}_{r}r_{x}) + \mathbb{K}^{0}(x,t,q,r)
\end{equation}

\noindent
Plugging this into $\eqref{Step2}$ gives us

\begin{eqnarray}
&& \mathbb{F}_{t} - i\mathbb{F}_{r}(gr^{2}q + (\upsilon-i\gamma)r) + i\mathbb{F}_{q}(gq^{2}r + (\upsilon+i\gamma)q) - if_{x}(\mathbb{F}_{q}q_{x} - \mathbb{F}_{r}r_{x}) - \mathbb{K}^{0}_{q}q_{x} - \mathbb{K}^{0}_{r}r_{x} \nonumber \\ 
&& - if(\mathbb{F}_{qx}q_{x} - \mathbb{F}_{rx}r_{x}) - if(\mathbb{F}_{qq}q_{x}^{2} - \mathbb{F}_{rr}r_{x}^{2}) - \mathbb{K}^{0}_{x} + ifq_{x}\left[\mathbb{F},\mathbb{F}_{q}\right] - ifr_{x}\left[\mathbb{F},\mathbb{F}_{r}\right] + \left[\mathbb{F},\mathbb{K}^{0}\right] = 0 \label{Step3}
\end{eqnarray}

Now since $\mathbb{F}$ and $\mathbb{K}^{0}$ do not depend on $q_{x}$ and $r_{x}$ we collect the coefficients of the $q_{x}^{2}$ and $r_{x}^{2}$ and equate them to zero. This now requires

\[ if\mathbb{F}_{rr} = 0 = -if\mathbb{F}_{qq} \]

\noindent
from which it follows via simple integration

\[ \mathbb{F} = \mathbb{X}_{1}(x,t) + \mathbb{X}_{2}(x,t)q + \mathbb{X}_{3}(x,t)r + \mathbb{X}_{4}(x,t)rq \]

\noindent
where the $\mathbb{X}_{i}$ are arbitrary matrices whose elements are functions of $x$ and $t$. Plugging this into $\eqref{Step3}$ we obtain

\begin{eqnarray}
&& \mathbb{X}_{1,t} + \mathbb{X}_{2,t}q + \mathbb{X}_{3,t}r + \mathbb{X}_{4,t}rq - i(\mathbb{X}_{3} + \mathbb{X}_{4}q)(gr^{2}q + (\upsilon-i\gamma)r) + i(\mathbb{X}_{2} + \mathbb{X}_{4}r)(gq^{2}r + (\upsilon+i\gamma)q) \nonumber \\
&& - if_{x}((\mathbb{X}_{2} + \mathbb{X}_{4}r)q_{x} - (\mathbb{X}_{3} + \mathbb{X}_{4}q)r_{x}) - \mathbb{K}^{0}_{q}q_{x} - \mathbb{K}^{0}_{r}r_{x} - if((\mathbb{X}_{2,x} + \mathbb{X}_{4,x}r)q_{x} - (\mathbb{X}_{3,x} + \mathbb{X}_{4,x}q)r_{x}) \nonumber \\
&& - \mathbb{K}^{0}_{x} + if[\mathbb{X}_{1},\mathbb{X}_{2}]q_{x} + if[\mathbb{X}_{3},\mathbb{X}_{2}]rq_{x} + if[\mathbb{X}_{4},\mathbb{X}_{2}]rqq_{x} + if[\mathbb{X}_{1},\mathbb{X}_{4}]rq_{x} + if[\mathbb{X}_{2},\mathbb{X}_{4}]rqq_{x} \nonumber \\
&& + if[\mathbb{X}_{3},\mathbb{X}_{4}]r^{2}q_{x} - if[\mathbb{X}_{1},\mathbb{X}_{3}]r_{x} - if[\mathbb{X}_{2},\mathbb{X}_{3}]qr_{x} - if[\mathbb{X}_{4},\mathbb{X}_{3}]qrr_{x} - if[\mathbb{X}_{1},\mathbb{X}_{4}]qr_{x} \nonumber \\
&& - if[\mathbb{X}_{2},\mathbb{X}_{4}]q^{2}r_{x} - if[\mathbb{X}_{3},\mathbb{X}_{4}]qrr_{x} + [\mathbb{X}_{1},\mathbb{K}^{0}] + [\mathbb{X}_{2},\mathbb{K}^{0}]q + [\mathbb{X}_{3},\mathbb{K}^{0}]r + [\mathbb{X}_{4},\mathbb{K}^{0}]qr = 0 \label{Step4a}
\end{eqnarray}

Noting the antisymmetry of the commutator we can further simplify this to

\begin{eqnarray}
&& \mathbb{X}_{1,t} + \mathbb{X}_{2,t}q + \mathbb{X}_{3,t}r + \mathbb{X}_{4,t}rq - i(\mathbb{X}_{3} + \mathbb{X}_{4}q)(gr^{2}q + (\upsilon-i\gamma)r) + i(\mathbb{X}_{2} + \mathbb{X}_{4}r)(gq^{2}r + (\upsilon+i\gamma)q) \nonumber \\
&& - if_{x}((\mathbb{X}_{2} + \mathbb{X}_{4}r)q_{x} - (\mathbb{X}_{3} + \mathbb{X}_{4}q)r_{x}) - \mathbb{K}^{0}_{q}q_{x} - \mathbb{K}^{0}_{r}r_{x} - if((\mathbb{X}_{2,x} + \mathbb{X}_{4,x}r)q_{x} - (\mathbb{X}_{3,x} + \mathbb{X}_{4,x}q)r_{x})  \nonumber \\
&& - \mathbb{K}^{0}_{x} + if[\mathbb{X}_{1},\mathbb{X}_{2}]q_{x} + if[\mathbb{X}_{3},\mathbb{X}_{2}]rq_{x} + if[\mathbb{X}_{1},\mathbb{X}_{4}]rq_{x} + if[\mathbb{X}_{3},\mathbb{X}_{4}]r^{2}q_{x} - if[\mathbb{X}_{1},\mathbb{X}_{3}]r_{x} \nonumber \\
&& - if[\mathbb{X}_{2},\mathbb{X}_{3}]qr_{x} - if[\mathbb{X}_{1},\mathbb{X}_{4}]qr_{x} - if[\mathbb{X}_{2},\mathbb{X}_{4}]q^{2}r_{x} = 0 \label{Step4b}
\end{eqnarray}

As before, since the $\mathbb{X}_{i}$ and $\mathbb{K}^{0}$ do not depend on $r_{x}$ or $q_{x}$ we equate the coefficients of the $q_{x}$ and $r_{x}$ terms to zero. Thus we require

\begin{eqnarray} 
&& -if_{x}(\mathbb{X}_{2} + \mathbb{X}_{4}r) - \mathbb{K}^{0}_{q} - if(\mathbb{X}_{2,x} + \mathbb{X}_{4,x}r) + if[\mathbb{X}_{1},\mathbb{X}_{2}] + ifr[\mathbb{X}_{1},\mathbb{X}_{4}] \nonumber \\
&& - ifr[\mathbb{X}_{2},\mathbb{X}_{3}] + ifr^{2}[\mathbb{X}_{3},\mathbb{X}_{4}] = 0 \label{FindKa} \\ 
&& if_{x}(\mathbb{X}_{3} + \mathbb{X}_{4}q) - \mathbb{K}^{0}_{r} + if(\mathbb{X}_{3,x} + \mathbb{X}_{4,x}q) - if[\mathbb{X}_{1},\mathbb{X}_{3}] - ifq[\mathbb{X}_{1},\mathbb{X}_{4}] \nonumber \\
&& - ifq[\mathbb{X}_{2},\mathbb{X}_{3}] - ifq^{2}[\mathbb{X}_{2},\mathbb{X}_{4}] = 0 \label{FindKb}
\end{eqnarray}

Upon trying to integrate this system one finds that the system is in fact inconsistent. Recall that given a system of PDEs
\[ \Psi_{q} = \xi(q,r), \ \ \Psi_{r} = \eta(q,r) \]
if we are to recover $\Psi$ we must satisfy a consistency condition. That is, we must have $\xi_{r} = \Psi_{qr} = \Psi_{rq} = \eta_{q}$. In $\eqref{FindKa}$ and $\eqref{FindKb}$ we have

\begin{eqnarray}
\xi(q,r) &=& -if_{x}(\mathbb{X}_{2} + \mathbb{X}_{4}r) - if(\mathbb{X}_{2,x} + \mathbb{X}_{4,x}r) + if[\mathbb{X}_{1},\mathbb{X}_{2}] + ifr[\mathbb{X}_{1},\mathbb{X}_{4}] \nonumber \\
&& - ifr[\mathbb{X}_{2},\mathbb{X}_{3}] + ifr^{2}[\mathbb{X}_{3},\mathbb{X}_{4}] = 0 \\ 
\eta(q,r) &=& if_{x}(\mathbb{X}_{3} + \mathbb{X}_{4}q) + if(\mathbb{X}_{3,x} + \mathbb{X}_{4,x}q) - if[\mathbb{X}_{1},\mathbb{X}_{3}] - ifq[\mathbb{X}_{1},\mathbb{X}_{4}] \nonumber \\
&& - ifq[\mathbb{X}_{2},\mathbb{X}_{3}] - ifq^{2}[\mathbb{X}_{2},\mathbb{X}_{4}] = 0
\end{eqnarray}

Thus the consistency condition ($\xi_{r} = \eta_{q}$) requires

\begin{eqnarray*}
&& -if_{x}\mathbb{X}_{4} - if\mathbb{X}_{4,x} + if[\mathbb{X}_{1},\mathbb{X}_{4}] - if[\mathbb{X}_{2},\mathbb{X}_{3}] + 2if[\mathbb{X}_{3},\mathbb{X}_{4}]r = if_{x}\mathbb{X}_{4} + if\mathbb{X}_{4,x} - if[\mathbb{X}_{1},\mathbb{X}_{4}] \\
&& - if[\mathbb{X}_{2},\mathbb{X}_{3}] - 2if[\mathbb{X}_{2},\mathbb{X}_{4}]q
\end{eqnarray*}

\noindent
But this means we must have

\begin{equation}
2if_{x}\mathbb{X}_{4} + 2if\mathbb{X}_{4,x} - 2if[\mathbb{X}_{1},\mathbb{X}_{4}] - 2if[\mathbb{X}_{3},\mathbb{X}_{4}](r+q) = 0
\end{equation}

One easy choice to make the system consistent, and for the purpose of demonstrating how this method can reproduce results previously obtained in the literature, is to set $\mathbb{X}_{4} = 0$. Thus the system becomes

\begin{eqnarray} 
\mathbb{K}^{0}_{q} &=& -if_{x}\mathbb{X}_{2} - if\mathbb{X}_{2,x} + if[\mathbb{X}_{1},\mathbb{X}_{2}] - ifr[\mathbb{X}_{2},\mathbb{X}_{3}] \label{Step5a} \\
\mathbb{K}^{0}_{r} &=& if_{x}\mathbb{X}_{3} + if\mathbb{X}_{3,x} - if[\mathbb{X}_{1},\mathbb{X}_{3}] - ifq[\mathbb{X}_{2},\mathbb{X}_{3}] \label{Step5b}
\end{eqnarray}

\noindent
Integrating the first equation with respect to $q$ we obtain

\[ \mathbb{K}^{0} = -if_{x}\mathbb{X}_{2}q - if\mathbb{X}_{2,x}q + if[\mathbb{X}_{1},\mathbb{X}_{2}]q - if[\mathbb{X}_{2},\mathbb{X}_{3}]rq + \mathbb{K}^{*}(x,t,r) \]

\noindent
Now differentiating this and mandating that it equal our previous expression for $K^{0}_{r}$ we find that $K^{*}$ must satisfy

\[ \mathbb{K}^{*}_{r} = if_{x}\mathbb{X}_{3} + if\mathbb{X}_{3,x} - if[\mathbb{X}_{1},\mathbb{X}_{3}]  \]

Integrating this expression with respect to $r$ we easily find

\[ \mathbb{K}^{*} = if_{x}\mathbb{X}_{3}r + if\mathbb{X}_{3,x}r - if[\mathbb{X}_{1},\mathbb{X}_{3}]r + \mathbb{X}_{0}(x,t) \]

\noindent
Now plugging this into our previous expression for $\mathbb{K}^{0}$ we have

\begin{equation}
\mathbb{K}^{0} = if_{x}(\mathbb{X}_{3}r - \mathbb{X}_{2}q) + if(\mathbb{X}_{3,x}r - \mathbb{X}_{2,x}q) + if[\mathbb{X}_{1},\mathbb{X}_{2}]q - if[\mathbb{X}_{1},\mathbb{X}_{3}]r - if[\mathbb{X}_{2},\mathbb{X}_{3}]qr + \mathbb{X}_{0}
\end{equation}

\noindent
Now plugging this into $\eqref{Step4b}$ we have

\begin{eqnarray}
&& \mathbb{X}_{1,t} + \mathbb{X}_{2,t}q + \mathbb{X}_{3,t}r - i\mathbb{X}_{3}(gr^{2}q + (\upsilon-i\gamma)r) + i\mathbb{X}_{2}(gq^{2}r + (\upsilon+i\gamma)q) - if_{xx}(\mathbb{X}_{3}r - \mathbb{X}_{2}q) \nonumber \\
&& - 2if_{x}(\mathbb{X}_{3,x}r - \mathbb{X}_{2,x}q) - if(\mathbb{X}_{3,xx}r - \mathbb{X}_{2,xx}q) - i(f[\mathbb{X}_{1},\mathbb{X}_{2}])_{x}q + i(f[\mathbb{X}_{1},\mathbb{X}_{3}])_{x}r \nonumber \\
&& - \mathbb{X}_{0,x} + if_{x}([\mathbb{X}_{1},\mathbb{X}_{3}]r - [\mathbb{X}_{1},\mathbb{X}_{2}]q) + if([\mathbb{X}_{1},\mathbb{X}_{3,x}]r - [\mathbb{X}_{1},\mathbb{X}_{2,x}]q) + if[\mathbb{X}_{1},[\mathbb{X}_{1},\mathbb{X}_{2}]]q \nonumber \\
&& - if[\mathbb{X}_{1},[\mathbb{X}_{1},\mathbb{X}_{3}]]r - if[\mathbb{X}_{1},[\mathbb{X}_{2},\mathbb{X}_{3}]]qr + [\mathbb{X}_{1},\mathbb{X}_{0}] + if_{x}[\mathbb{X}_{2},\mathbb{X}_{3}]qr + i(f[\mathbb{X}_{2},\mathbb{X}_{3}])_{x}qr \nonumber \\
&& + if[\mathbb{X}_{2},[\mathbb{X}_{1},\mathbb{X}_{2}]]q^{2} - if[\mathbb{X}_{2},[\mathbb{X}_{1},\mathbb{X}_{3}]]qr - if[\mathbb{X}_{2},[\mathbb{X}_{2},\mathbb{X}_{3}]]q^{2}r + [\mathbb{X}_{2},\mathbb{X}_{0}]q - if_{x}[\mathbb{X}_{3},\mathbb{X}_{2}]rq \nonumber \\
&& + if([\mathbb{X}_{3},\mathbb{X}_{3,x}]r^{2} - [\mathbb{X}_{3},\mathbb{X}_{2,x}]rq) + if[\mathbb{X}_{3},[\mathbb{X}_{1},\mathbb{X}_{2}]]rq - if[\mathbb{X}_{3},[\mathbb{X}_{1},\mathbb{X}_{3}]]r^{2} - if[\mathbb{X}_{3},[\mathbb{X}_{2},\mathbb{X}_{3}]]qr^{2} \nonumber \\
&& + if([\mathbb{X}_{2},\mathbb{X}_{3,x}]qr - [\mathbb{X}_{2},\mathbb{X}_{2,x}]q^{2}) + [\mathbb{X}_{3},\mathbb{X}_{0}]r = 0 \label{Step6}
\end{eqnarray}

Since the $\mathbb{X}_{i}$ are independent of $r$ and $q$ we equate the coefficients of the different powers of $r$ and $q$ to zero and thus obtain the following constraints:

\begin{eqnarray}
O(1) &:& \mathbb{X}_{1,t} - \mathbb{X}_{0,x} + \left[\mathbb{X}_{1},\mathbb{X}_{0}\right] = 0 \\
O(q) &:& \mathbb{X}_{2,t} + i\mathbb{X}_{2}(\upsilon + i\gamma) + i(f\mathbb{X}_{2})_{xx} - i(f[\mathbb{X}_{1},\mathbb{X}_{2}])_{x} - if_{x}[\mathbb{X}_{1},\mathbb{X}_{2}] - if[\mathbb{X}_{1},\mathbb{X}_{2,x}] \nonumber \\
&& + if[\mathbb{X}_{1},[\mathbb{X}_{1},\mathbb{X}_{2}]] + [\mathbb{X}_{2},\mathbb{X}_{0}] = 0 \\
O(r) &:& \mathbb{X}_{3,t} - i\mathbb{X}_{3}(\upsilon - i\gamma) - i(f\mathbb{X}_{3})_{xx} + i(f[\mathbb{X}_{1},\mathbb{X}_{3}])_{x} + if_{x}[\mathbb{X}_{1},\mathbb{X}_{3}] + if[\mathbb{X}_{1},\mathbb{X}_{3,x}] \nonumber \\
&& - if[\mathbb{X}_{1},[\mathbb{X}_{1},\mathbb{X}_{3}]] + [\mathbb{X}_{3},\mathbb{X}_{0}] = 0 \\
O(qr) &:& 2i(f[\mathbb{X}_{2},\mathbb{X}_{3}])_{x} - if[\mathbb{X}_{1},[\mathbb{X}_{2},\mathbb{X}_{3}]] + if_{x}[\mathbb{X}_{2},\mathbb{X}_{3}] - if[\mathbb{X}_{2},[\mathbb{X}_{1},\mathbb{X}_{3}]] \nonumber \\
&& + if[\mathbb{X}_{3},[\mathbb{X}_{1},\mathbb{X}_{2}]] = 0 \\
O(q^{2}) &:& if[\mathbb{X}_{2},\mathbb{X}_{2,x}] - if[\mathbb{X}_{2},[\mathbb{X}_{1},\mathbb{X}_{2}]] = 0 \label{Satisf1} \\
O(r^{2}) &:& if[\mathbb{X}_{3},\mathbb{X}_{3,x}] - if[\mathbb{X}_{3},[\mathbb{X}_{1},\mathbb{X}_{3}]] = 0 \label{Satisf2} \\
O(q^{2}r) &:& ig\mathbb{X}_{2} - if[\mathbb{X}_{2},[\mathbb{X}_{2},\mathbb{X}_{3}]] = 0 \\
O(r^{2}q) &:& ig\mathbb{X}_{3} + if[\mathbb{X}_{3},[\mathbb{X}_{2},\mathbb{X}_{3}]] = 0
\end{eqnarray}

These equations collectively determine the conditions for integrability of the system. Note that in general, as with the standard Estabrook-Wahlquist method, the solution to the above system is not unique. Provided we can find representations for the $\mathbb{X}_{i}$ and thus reduce the system down to an integrability condition on the coefficients we will obtain our Lax pair $\mathbb{F}$ and $\mathbb{G}$. We will now show how to reproduce the results given in Khawaja's paper. Let us consider Khawaja's choices, thus

\begin{equation}
\mathbb{X}_{0} = \begin{bmatrix}
g_{1} & 0 \\
0 & g_{13}
\end{bmatrix}, \ \ \mathbb{X}_{1} = \begin{bmatrix}
f_{1} & 0 \\
0 & f_{7}
\end{bmatrix}, \ \ \mathbb{X}_{2} = \begin{bmatrix}
0 & ip_{1} \\
0 & 0
\end{bmatrix}, \ \ \mathbb{X}_{3} = \begin{bmatrix}
0 & 0 \\
-ip_{2} & 0
\end{bmatrix}
\end{equation}

Plugging this into our integrability conditions yields

\begin{eqnarray} 
O(1) &:& f_{1t} - g_{1x} = 0 \label{KC1} \\ 
O(1) &:& f_{7t} - g_{13x} = 0 \label{KC2} \\ 
O(q) &:& ip_{1t} - ip_{1}(g_{1} - g_{13} - i\upsilon + \gamma) - (fp_{1})_{xx} + 2(f_{1}-f_{7})(p_{1}f)_{x} \nonumber \\ 
&& -fp_{1}(f_{1}-f_{7})^{2} + fp_{1}(f_{1}-f_{7})_{x} = 0 \label{KC3} \\ 
O(r) &:& ip_{2t} + ip_{2}(g_{1} - g_{13} - i\upsilon - \gamma) + (fp_{2})_{xx} + 2(f_{1}-f_{7})(fp_{2})_{x} \nonumber \\ 
&& + (f_{1}-f_{7})^{2}fp_{2} + fp_{2}(f_{1}-f_{7})_{x} = 0 \label{KC4} \\ 
O(qr) &:& f_{x}p_{1}p_{2} + 2(fp_{1}p_{2})_{x} = 0 \label{KC5} \\ 
O(q^{2}r) \ \mbox{and} \ O(r^{2}q) &:& g + 2fp_{1}p_{2} = 0 \label{KC6}
\end{eqnarray}

Note that $\eqref{Satisf1}$ and $\eqref{Satisf2}$ were identically satisfied. {\it In Khawaja's paper we see $\eqref{KC1},\eqref{KC2},\eqref{KC5}$ and $\eqref{KC6}$ given exactly.} To see that the other conditions are equivalent we note that in his paper he had the additional determining equations

\begin{eqnarray} 
&& (fp_{1})_{x} - fp_{1}(f_{1}-f_{7}) - g_{6} = 0 \label{EKC1} \\ 
&& (fp_{2})_{x} + fp_{2}(f_{1}-f_{7}) - g_{10} = 0 \label{EKC2} \\ 
&& g_{6}(f_{1}-f_{7}) - ip_{1}(g_{1} - g_{13} - i\upsilon + \gamma) - g_{6x} + ip_{1t} = 0 \label{EKC3} \\ 
&& g_{10}(f_{1}-f_{7}) + ip_{2}(g_{1} - g_{13} - i\upsilon - \gamma) + g_{10x} + ip_{2t} = 0 \label{EKC4}
\end{eqnarray}

We begin by solving $\eqref{EKC1}$ and $\eqref{EKC2}$ for $g_{6}$ and $g_{10}$, respectively. Now plugging $g_{6}$ into $\eqref{EKC3}$ and $g_{10}$ into $\eqref{EKC4}$ we obtain

\begin{eqnarray}
&& 2(fp_{1})_{x}(f_{1}-f_{7}) - fp_{1}(f_{1}-f_{7})^{2} - ip_{1}(g_{1} - g_{13} - i\upsilon + \gamma) - (fp_{1})_{xx} \nonumber \\
&& + fp_{1}(f_{1}-f_{7})_{x} + ip_{1t} = 0 \\
&& (fp_{2})_{x}(f_{1}-f_{7}) + fp_{2}(f_{1}-f_{7})^{2} + ip_{2}(g_{1} - g_{13} - i\upsilon - \gamma) + (fp_{2})_{xx} \nonumber \\
&& + fp_{2}(f_{1}-f_{7})_{x} + ip_{2t} = 0
\end{eqnarray}

\noindent
which is exactly $\eqref{KC3}$ and $\eqref{KC4}$. 

To demonstrate how one can obtain the results for a constant coefficient version of the NLPDE we now take the $\mathbb{X}_{i}$ to be constant matrices (as in the standard method) and consider $f = -\frac{1}{2}$, $g = 1$, and $\upsilon = \gamma = 0$. This reduces $\eqref{Khawaja}$ to the standard NLS

\[ iq_{t} - \frac{1}{2}q_{xx} + |q|^{2}q = 0 \]

The previous conditions reduce to

\begin{eqnarray}
O(1) &:& \left[\mathbb{X}_{1},\mathbb{X}_{0}\right] = 0 \\
O(q) &:&  -i\frac{1}{2}[\mathbb{X}_{1},[\mathbb{X}_{1},\mathbb{X}_{2}]] + [\mathbb{X}_{2},\mathbb{X}_{0}] = 0 \\
O(r) &:& i\frac{1}{2}[\mathbb{X}_{1},[\mathbb{X}_{1},\mathbb{X}_{3}]] + [\mathbb{X}_{3},\mathbb{X}_{0}] = 0 \\
O(qr) &:& [\mathbb{X}_{1},[\mathbb{X}_{2},\mathbb{X}_{3}]] + [\mathbb{X}_{2},[\mathbb{X}_{1},\mathbb{X}_{3}]] - [\mathbb{X}_{3},[\mathbb{X}_{1},\mathbb{X}_{2}]] = 0 \\
O(q^{2}) &:& [\mathbb{X}_{2},[\mathbb{X}_{1},\mathbb{X}_{2}]] = 0 \\
O(r^{2}) &:& [\mathbb{X}_{3},[\mathbb{X}_{1},\mathbb{X}_{3}]] = 0 \\
O(q^{2}r) &:& \mathbb{X}_{2} - \frac{1}{2}[\mathbb{X}_{2},[\mathbb{X}_{2},\mathbb{X}_{3}]] = 0 \\
O(r^{2}q) &:& \mathbb{X}_{3} + \frac{1}{2}[\mathbb{X}_{3},[\mathbb{X}_{2},\mathbb{X}_{3}]] = 0
\end{eqnarray}

{\it One can verify that a solution to this system is given by}

\begin{equation}
\mathbb{X}_{0} = \begin{bmatrix}
i\lambda^{2} & 0 \\
0 & -i\lambda^{2}
\end{bmatrix}, \ \ \mathbb{X}_{1} = \begin{bmatrix}
i\lambda & 0 \\
0 & -i\lambda
\end{bmatrix}, \ \ \mathbb{X}_{2} = \begin{bmatrix}
0 & i \\
0 & 0
\end{bmatrix}, \ \ \mathbb{X}_{3} = \begin{bmatrix}
0 & 0 \\
i & 0
\end{bmatrix}
\end{equation}

{\it This solution is exactly that which one derives in the AKNS scheme. Therefore as one can see, the system of algebraic equations for a set of generators which is derived as a necessary condition for the Lax integrability of an NLPDE are merely a reduction of a larger system of (possibly nonlinear) PDEs which represent a generalization of the NLPDE.}

\section{The Generalized Fifth-Order Korteweg-deVries (KdV) Equation}

As a second example consider the generalized KDV equation

\begin{equation} \label{KDV}
u_{t} + a_{1}uu_{xxx} + a_{2}u_{x}u_{xx} + a_{3}u^{2}u_{x} + a_{4}uu_{x} + a_{5}u_{xxx} + a_{6}u_{xxxxx} + a_{7}u + a_{8}u_{x} = 0
\end{equation}

\noindent
where $a_{1-8}$ are arbitrary functions of $x$ and $t$. As with the last example, we will go through the procedure outlined earlier in the paper and show how one can obtain the results previously obtained for the constant coefficient cases. Running through the standard procedure we let $\mathbb{F} = \mathbb{F}(x,t,u)$ and $\mathbb{G} = \mathbb{G}(x,t,u,u_{x},u_{xx},u_{xxx},u_{xxxx})$. Plugging this into $\eqref{ZCC}$ we obtain

\begin{equation}
\mathbb{F}_{t} + \mathbb{F}_{u}u_{t} - \mathbb{G}_{x} - \mathbb{G}_{u}u_{x} - \mathbb{G}_{u_{x}}u_{xx} - \mathbb{G}_{u_{xx}}u_{xxx} - \mathbb{G}_{u_{xxx}}u_{xxxx} - \mathbb{G}_{u_{xxxx}}u_{xxxxx} + [\mathbb{F},\mathbb{G}] = 0
\end{equation}

\noindent
Next, substituting $\eqref{KDV}$ into this expression in order to eliminate the $u_{t}$ yields

\begin{eqnarray} \nonumber
&& \mathbb{F}_{t} - \mathbb{F}_{u}\left(a_{1}uu_{xxx} + a_{2}u_{x}u_{xx} + a_{3}u^{2}u_{x} + a_{4}uu_{x} + a_{5}u_{xxx} + a_{6}u_{xxxxx} + a_{7}u + a_{8}u_{x}\right) - \mathbb{G}_{x} \\ \label{KDVstep1}
&& - \mathbb{G}_{u}u_{x} - \mathbb{G}_{u_{x}}u_{xx} - \mathbb{G}_{u_{xx}}u_{xxx} - \mathbb{G}_{u_{xxx}}u_{xxxx} - \mathbb{G}_{u_{xxxx}}u_{xxxxx} + [\mathbb{F},\mathbb{G}] = 0
\end{eqnarray}

Since $\mathbb{F}$ and $\mathbb{G}$ do not depend on $u_{xxxxx}$ we can equate the coefficient of the $u_{xxxxx}$ term to zero. This requires that we must have

\[ \mathbb{G}_{u_{xxxx}} + a_{6}\mathbb{F}_{u} = 0 \Rightarrow \mathbb{G} = -a_{6}\mathbb{F}_{u}u_{xxxx} + \mathbb{K}^{0}(x,t,u,u_{x},u_{xx},u_{xxx}) \]

\noindent
Now updating $\eqref{KDVstep1}$ we obtain

\begin{eqnarray}
&& \mathbb{F}_{t} - \mathbb{F}_{u}\left(a_{1}uu_{xxx} + a_{2}u_{x}u_{xx} + a_{3}u^{2}u_{x} + a_{4}uu_{x} + a_{5}u_{xxx} + a_{7}u + a_{8}u_{x}\right) + a_{6x}\mathbb{F}_{u}u_{xxxx} \nonumber \\
&&  + a_{6}\mathbb{F}_{xu}u_{xxxx} - \mathbb{K}^{0}_{x} - \mathbb{K}^{0}_{u}u_{x} - \mathbb{K}^{0}_{u_{x}}u_{xx} - \mathbb{K}^{0}_{u_{xx}}u_{xxx} - \mathbb{K}^{0}_{u_{xxx}}u_{xxxx} + a_{6}\mathbb{F}_{uu}u_{x}u_{xxxx} \nonumber \\ 
&& - [\mathbb{F},\mathbb{F}_{u}]a_{6}u_{xxxx} + [\mathbb{F},\mathbb{K}^{0}] = 0 \label{KDVstep2}
\end{eqnarray}

Since $\mathbb{F}$ and $\mathbb{K}^{0}$ do not depend on $u_{xxxx}$ we can equate the coefficient of the $u_{xxxx}$ term to zero. This requires that we have

\begin{equation}
a_{6x}\mathbb{F}_{u} + a_{6}\mathbb{F}_{xu} + a_{6}\mathbb{F}_{uu}u_{x}- \mathbb{K}^{0}_{u_{xxx}} - [\mathbb{F},\mathbb{F}_{u}]a_{6} = 0
\end{equation}

\noindent
Thus, integrating with respect to $u_{xxx}$ and solving for $\mathbb{K}^{0}$ we have

\begin{equation}
\mathbb{K}^{0} = a_{6x}\mathbb{F}_{u}u_{xxx} + a_{6}\mathbb{F}_{xu}u_{xxx} + a_{6}\mathbb{F}_{uu}u_{x}u_{xxx} - [\mathbb{F},\mathbb{F}_{u}]a_{6}u_{xxx} + \mathbb{K}^{1}(x,t,u,u_{x},u_{xx})
\end{equation}

\noindent
Now we update $\eqref{KDVstep2}$ by plugging in our expression for $\mathbb{K}^{1}$ to obtain

\begin{eqnarray} 
&& \mathbb{F}_{t} - \mathbb{F}_{u}\left(a_{1}uu_{xxx} + a_{2}u_{x}u_{xx} + a_{3}u^{2}u_{x} + a_{4}uu_{x} + a_{5}u_{xxx} + a_{7}u + a_{8}u_{x}\right) - a_{6xx}\mathbb{F}_{u}u_{xxx} \nonumber \\
&& - 2a_{6x}\mathbb{F}_{xu}u_{xxx} - a_{6}\mathbb{F}_{xxu}u_{xxx} - a_{6x}\mathbb{F}_{uu}u_{x}u_{xxx} - a_{6}\mathbb{F}_{xuu}u_{x}u_{xxx} + [\mathbb{F}_{x},\mathbb{F}_{u}]a_{6}u_{xxx} \nonumber \\
&& + [\mathbb{F},\mathbb{F}_{xu}]a_{6}u_{xxx} + [\mathbb{F},\mathbb{F}_{u}]a_{6x}u_{xxx} - \mathbb{K}^{1}_{x} - a_{6x}\mathbb{F}_{uu}u_{x}u_{xxx} - a_{6}\mathbb{F}_{xuu}u_{x}u_{xxx} \nonumber \\
&& - a_{6}\mathbb{F}_{uuu}u_{x}^{2}u_{xxx} + [\mathbb{F},\mathbb{F}_{uu}]a_{6}u_{x}u_{xxx} - \mathbb{K}^{1}_{u}u_{x} - a_{6}\mathbb{F}_{uu}u_{xx}u_{xxx} - \mathbb{K}^{1}_{u_{x}}u_{xx} - \mathbb{K}^{1}_{u_{xx}}u_{xxx} \nonumber \\ 
&&  + a_{6x}[\mathbb{F},\mathbb{F}_{u}]u_{xxx} + a_{6}[\mathbb{F},\mathbb{F}_{xu}]u_{xxx} + a_{6}[\mathbb{F},\mathbb{F}_{uu}]u_{x}u_{xxx} - [\mathbb{F},[\mathbb{F},\mathbb{F}_{u}]]a_{6}u_{xxx} + [\mathbb{F},\mathbb{K}^{1}] = 0 \label{KDV3}
\end{eqnarray}

Since $\mathbb{F}$ and $\mathbb{K}^{1}$ do not depend on $u_{xxx}$ we can equate the coefficient of the $u_{xxx}$ term to zero. This requires that we have

\begin{eqnarray} 
&& - \mathbb{F}_{u}(a_{1}u + a_{5}) - a_{6xx}\mathbb{F}_{u} - 2a_{6x}\mathbb{F}_{xu} - a_{6}\mathbb{F}_{xxu} - a_{6x}\mathbb{F}_{uu}u_{x} - a_{6}\mathbb{F}_{xuu}u_{x} \nonumber \\
&& + [\mathbb{F}_{x},\mathbb{F}_{u}]a_{6} + [\mathbb{F},\mathbb{F}_{xu}]a_{6} + [\mathbb{F},\mathbb{F}_{u}]a_{6x} - a_{6x}\mathbb{F}_{uu}u_{x} - a_{6}\mathbb{F}_{xuu}u_{x} - a_{6}\mathbb{F}_{uuu}u_{x}^{2} \nonumber \\
&& + [\mathbb{F},\mathbb{F}_{uu}]a_{6}u_{x} - a_{6}\mathbb{F}_{uu}u_{xx} - \mathbb{K}^{1}_{u_{xx}} + a_{6x}[\mathbb{F},\mathbb{F}_{u}] + a_{6}[\mathbb{F},\mathbb{F}_{xu}] + a_{6}[\mathbb{F},\mathbb{F}_{uu}]u_{x} \nonumber \\ 
&& - [\mathbb{F},[\mathbb{F},\mathbb{F}_{u}]]a_{6} = 0
\end{eqnarray}

\noindent
Integrating with respect to $u_{xx}$ and solving for $\mathbb{K}^{1}$ and collecting like terms we have

\begin{eqnarray}
\mathbb{K}^{1} &=& -\mathbb{F}_{u}(a_{1}u + a_{5})u_{xx} - (a_{6}\mathbb{F}_{u})_{xx}u_{xx} - 2(a_{6}\mathbb{F}_{uu})_{x}u_{x}u_{xx} + 2(a_{6}[\mathbb{F},\mathbb{F}_{u}])_{x}u_{xx} \nonumber \\
&& - a_{6}\mathbb{F}_{uuu}u_{x}^{2}u_{xx} + 2a_{6}[\mathbb{F},\mathbb{F}_{uu}]u_{x}u_{xx} - \frac{1}{2}a_{6}\mathbb{F}_{uu}u_{xx}^{2} - a_{6}[\mathbb{F}_{x},\mathbb{F}_{u}]u_{xx} \nonumber \\
&& - a_{6}[\mathbb{F},[\mathbb{F},\mathbb{F}_{u}]]u_{xx} + \mathbb{K}^{2}(x,t,u,u_{x}) \label{KDV4}
\end{eqnarray}

Plugging $\eqref{KDV4}$ into $\eqref{KDV3}$ and simplifying a little bit we obtain

\begin{eqnarray} 
&& \mathbb{F}_{t} - \mathbb{F}_{u}(a_{2}u_{x}u_{xx} + a_{3}u^{2}u_{x} + a_{4}uu_{x} + a_{7}u + a_{8}u_{x}) + (a_{1}\mathbb{F}_{u})_{x}uu_{xx} + (a_{5}\mathbb{F}_{u})_{x}u_{xx} \nonumber \\
&& + (a_{6}\mathbb{F}_{u})_{xxx}u_{xx} + 2(a_{6}\mathbb{F}_{uu})_{xx}u_{x}u_{xx} - (a_{6}[\mathbb{F},\mathbb{F}_{u}])_{xx}u_{xx} + (a_{6}\mathbb{F}_{uuu})_{x}u_{x}^{2}u_{xx} \nonumber \\
&& + \frac{1}{2}(a_{6}\mathbb{F}_{uu})_{x}u_{xx}^{2}  - ([\mathbb{F},(a_{6}\mathbb{F}_{u})_{x}])_{x}u_{xx} + (a_{6}[\mathbb{F},[\mathbb{F},\mathbb{F}_{u}]])_{x}u_{xx} - \mathbb{K}^{2}_{x} + \mathbb{F}_{uu}a_{1}uu_{x}u_{xx} \nonumber \\
&& + \mathbb{F}_{u}a_{1}u_{x}u_{xx} + \mathbb{F}_{uu}a_{5}u_{x}u_{xx} + (a_{6}\mathbb{F}_{uu})_{xx}u_{x}u_{xx} + 2(a_{6}\mathbb{F}_{uuu})_{x}u_{x}^{2}u_{xx} + a_{6}\mathbb{F}_{uuuu}u_{x}^{3}u_{xx} \nonumber \\
&& - a_{6}[\mathbb{F}_{u},\mathbb{F}_{uu}]u_{x}^{2}u_{xx} - a_{6}[\mathbb{F},\mathbb{F}_{uuu}]u_{x}^{2}u_{xx} + \frac{5}{2}a_{6}\mathbb{F}_{uuu}u_{xx}^{2}u_{x} - [\mathbb{F}_{u},(a_{6}\mathbb{F}_{u})_{x}]u_{x}u_{xx} \nonumber \\
&& - 2[\mathbb{F},(a_{6}\mathbb{F}_{uu})_{x}]u_{x}u_{xx} - a_{6}[\mathbb{F}_{u},\mathbb{F}_{uu}]u_{x}^{2}u_{xx} - a_{6}[\mathbb{F},\mathbb{F}_{uuu}]u_{x}^{2}u_{xx} + a_{6}[\mathbb{F}_{u},[\mathbb{F},\mathbb{F}_{u}]]u_{x}u_{xx} \nonumber \\
&& + a_{6}[\mathbb{F},[\mathbb{F},\mathbb{F}_{uu}]]u_{x}u_{xx} - \mathbb{K}^{2}_{u}u_{x} + 2(a_{6}\mathbb{F}_{uu})_{x}u_{xx}^{2} - \frac{3}{2}a_{6}[\mathbb{F},\mathbb{F}_{uu}]u_{xx}^{2} - \mathbb{K}^{2}_{u_{x}}u_{xx} - a_{1}[\mathbb{F},\mathbb{F}_{u}]uu_{xx} \nonumber \\
&& - a_{5}[\mathbb{F},\mathbb{F}_{u}]u_{xx} - [\mathbb{F},(a_{6}\mathbb{F}_{u})_{xx}]u_{xx} - [\mathbb{F},(a_{6}\mathbb{F}_{uu})_{x}]u_{x}u_{xx} + [\mathbb{F},(a_{6}[\mathbb{F},\mathbb{F}_{u}])_{x}]u_{xx} \nonumber \\
&& - a_{6}[\mathbb{F},\mathbb{F}_{uuu}]u_{x}^{2}u_{xx} + 2a_{6}[\mathbb{F},[\mathbb{F},\mathbb{F}_{uu}]]u_{x}u_{xx} + [\mathbb{F},[\mathbb{F},(a_{6}\mathbb{F}_{u})_{x}]]u_{xx} - 3(a_{6}[\mathbb{F},\mathbb{F}_{uu}])_{x}u_{x}u_{xx} \nonumber \\
&& - a_{6}[\mathbb{F},[\mathbb{F},[\mathbb{F},\mathbb{F}_{u}]]]u_{xx} + [\mathbb{F},\mathbb{K}^{2}] = 0 \label{KDV5}
\end{eqnarray}

Now, since $\mathbb{K}^{2}$ and $\mathbb{F}$ do not depend on $u_{xx}$ we can start by setting the coefficients of the $u_{xx}^{2}$ and the $u_{xx}$ terms to zero. Note the difference here that we have multiple powers of $u_{xx}$ present in the $\eqref{KDV5}$. Setting the $O(u_{xx}^{2})$ term to zero requires

\begin{equation}
\frac{3}{2}(a_{6}\mathbb{F}_{uu})_{x} + \frac{5}{2}a_{6}\mathbb{F}_{uuu}u_{x} - \frac{3}{2}a_{6}[\mathbb{F},\mathbb{F}_{uu}]= 0 \label{DetermF}
\end{equation}

\noindent
Since $\mathbb{F}$ does not depend on $u_{x}$ we must have that the coefficient of the $u_{x}$ term in this previous expression is zero. This is equivalent to

\[ \mathbb{F}_{uuu} = 0 \Rightarrow \mathbb{F} = \mathbb{X}_{1}(x,t) + \mathbb{X}_{2}(x,t)u + \frac{1}{2}\mathbb{X}_{3}(x,t)u^{2} \]

\noindent
Plugging this into $\eqref{DetermF}$ we obtain

\begin{equation}
3(a_{6}\mathbb{X}_{3})_{x} - 3a_{6}([\mathbb{X}_{1},\mathbb{X}_{3}] + [\mathbb{X}_{2},\mathbb{X}_{3}]u) = 0 \label{CondX3}
\end{equation}

\noindent
Now since the $\mathbb{X}_{i}$ do not depend on $u$ we can set the coefficient of the $u$ to zero. That is, we require that $\mathbb{X}_{2}$ and $\mathbb{X}_{3}$ commute. We find now that $\eqref{CondX3}$ reduces to the condition

\begin{equation}
(a_{6}\mathbb{X}_{3})_{x} - a_{6}[\mathbb{X}_{1},\mathbb{X}_{3}] = 0 \label{CondX3final}
\end{equation}

\noindent
For ease of computation and in order to immediately satisfy $\eqref{CondX3final}$ we set $\mathbb{X}_{3} = 0$. Plugging into $\eqref{KDV5}$ our expression for $\mathbb{F}$ we obtain

\begin{eqnarray} 
&& \mathbb{X}_{1,t} + \mathbb{X}_{2,t}u - \mathbb{X}_{2}(a_{2}u_{x}u_{xx} + a_{3}u^{2}u_{x} + a_{4}uu_{x} + a_{7}u + a_{8}u_{x}) + (a_{1}\mathbb{X}_{2})_{x}uu_{xx} + (a_{5}\mathbb{X}_{2})_{x}u_{xx} \nonumber \\
&& + (a_{6}\mathbb{X}_{2})_{xxx}u_{xx} - (a_{6}[\mathbb{X}_{1},\mathbb{X}_{2}])_{xx}u_{xx} - \mathbb{K}^{2}_{x} + \mathbb{X}_{2}a_{1}u_{x}u_{xx} - [\mathbb{X}_{2},(a_{6}\mathbb{X}_{2})_{x}]u_{x}u_{xx} \nonumber \\
&& - ([\mathbb{X}_{1},(a_{6}\mathbb{X}_{2})_{x}])_{x}u_{xx} - ([\mathbb{X}_{2},(a_{6}\mathbb{X}_{2})_{x}])_{x}uu_{xx} + (a_{6}[\mathbb{X}_{1},[\mathbb{X}_{1},\mathbb{X}_{2}]])_{x}u_{xx} + (a_{6}[\mathbb{X}_{2},[\mathbb{X}_{1},\mathbb{X}_{2}]])_{x}uu_{xx} \nonumber \\
&& - \mathbb{K}^{2}_{u}u_{x} - \mathbb{K}^{2}_{u_{x}}u_{xx} - a_{1}[\mathbb{X}_{1},\mathbb{X}_{2}]uu_{xx} + a_{6}[\mathbb{X}_{2},[\mathbb{X}_{1},\mathbb{X}_{2}]]u_{x}u_{xx} + [\mathbb{X}_{2},(a_{6}[\mathbb{X}_{1},\mathbb{X}_{2}])_{x}]uu_{xx} \nonumber \\
&& - a_{5}[\mathbb{X}_{1},\mathbb{X}_{2}]u_{xx} - [\mathbb{X}_{1},(a_{6}\mathbb{X}_{2})_{xx}]u_{xx} - [\mathbb{X}_{2},(a_{6}\mathbb{X}_{2})_{xx}]uu_{xx} + [\mathbb{X}_{1},(a_{6}[\mathbb{X}_{1},\mathbb{X}_{2}])_{x}]u_{xx} \nonumber \\
&& + [\mathbb{X}_{1},[\mathbb{X}_{1},(a_{6}\mathbb{X}_{2})_{x}]]u_{xx} + [\mathbb{X}_{1},[\mathbb{X}_{2},(a_{6}\mathbb{X}_{2})_{x}]]uu_{xx} + [\mathbb{X}_{2},[\mathbb{X}_{1},(a_{6}\mathbb{X}_{2})_{x}]]uu_{xx} + [\mathbb{X}_{1},\mathbb{K}^{2}] \nonumber \\
&& - a_{6}[\mathbb{X}_{1},[\mathbb{X}_{1},[\mathbb{X}_{1},\mathbb{X}_{2}]]]u_{xx} - a_{6}[\mathbb{X}_{1},[\mathbb{X}_{2},[\mathbb{X}_{1},\mathbb{X}_{2}]]]uu_{xx} - a_{6}[\mathbb{X}_{2},[\mathbb{X}_{1},[\mathbb{X}_{1},\mathbb{X}_{2}]]]uu_{xx} \nonumber \\
&& + [\mathbb{X}_{2},[\mathbb{X}_{2},(a_{6}\mathbb{X}_{2})_{x}]]u^{2}u_{xx} - a_{6}[\mathbb{X}_{2},[\mathbb{X}_{2},[\mathbb{X}_{1},\mathbb{X}_{2}]]]u^{2}u_{xx} + [\mathbb{X}_{2},\mathbb{K}^{2}]u = 0 \label{KDV6}
\end{eqnarray}

Now again using the fact that the $\mathbb{X}_{i}$ and $\mathbb{K}^{2}$ do not depend on $u_{xx}$ we can set the coefficient of the $u_{xx}$ term in $\eqref{KDV6}$ to zero. This requires

\begin{eqnarray}
&& (a_{6}\mathbb{X}_{2})_{xxx} - (a_{6}[\mathbb{X}_{1},\mathbb{X}_{2}])_{xx} + a_{1}\mathbb{X}_{2}u_{x} - [\mathbb{X}_{2},(a_{6}\mathbb{X}_{2})_{x}]u_{x} - a_{2}\mathbb{X}_{2}u_{x} \nonumber \\
&& - ([\mathbb{X}_{1},(a_{6}\mathbb{X}_{2})_{x}])_{x} - ([\mathbb{X}_{2},(a_{6}\mathbb{X}_{2})_{x}])_{x}u + (a_{6}[\mathbb{X}_{1},[\mathbb{X}_{1},\mathbb{X}_{2}]])_{x} + (a_{6}[\mathbb{X}_{2},[\mathbb{X}_{1},\mathbb{X}_{2}]])_{x}u \nonumber \\
&& - \mathbb{K}^{2}_{u_{x}} - a_{1}[\mathbb{X}_{1},\mathbb{X}_{2}]u + a_{6}[\mathbb{X}_{2},[\mathbb{X}_{1},\mathbb{X}_{2}]]u_{x} + [\mathbb{X}_{2},(a_{6}[\mathbb{X}_{1},\mathbb{X}_{2}])_{x}]u + (a_{5}\mathbb{X}_{2})_{x} \nonumber \\
&& - a_{5}[\mathbb{X}_{1},\mathbb{X}_{2}] - [\mathbb{X}_{1},(a_{6}\mathbb{X}_{2})_{xx}] - [\mathbb{X}_{2},(a_{6}\mathbb{X}_{2})_{xx}]u + [\mathbb{X}_{1},(a_{6}[\mathbb{X}_{1},\mathbb{X}_{2}])_{x}] \nonumber \\
&& + [\mathbb{X}_{1},[\mathbb{X}_{1},(a_{6}\mathbb{X}_{2})_{x}]] + [\mathbb{X}_{1},[\mathbb{X}_{2},(a_{6}\mathbb{X}_{2})_{x}]]u + [\mathbb{X}_{2},[\mathbb{X}_{1},(a_{6}\mathbb{X}_{2})_{x}]]u + (a_{1}\mathbb{X}_{2})_{x}u \nonumber \\
&& - a_{6}[\mathbb{X}_{1},[\mathbb{X}_{1},[\mathbb{X}_{1},\mathbb{X}_{2}]]] - a_{6}[\mathbb{X}_{1},[\mathbb{X}_{2},[\mathbb{X}_{1},\mathbb{X}_{2}]]]u - a_{6}[\mathbb{X}_{2},[\mathbb{X}_{1},[\mathbb{X}_{1},\mathbb{X}_{2}]]]u \nonumber \\
&& + [\mathbb{X}_{2},[\mathbb{X}_{2},(a_{6}\mathbb{X}_{2})_{x}]]u^{2} - a_{6}[\mathbb{X}_{2},[\mathbb{X}_{2},[\mathbb{X}_{1},\mathbb{X}_{2}]]]u^{2} = 0 \label{KDV7}
\end{eqnarray}

\noindent
Thus integrating with respect to $u_{x}$ and solving for $\mathbb{K}^{2}$ we have

\begin{eqnarray} 
\mathbb{K}^{2} &=& (a_{6}\mathbb{X}_{2})_{xxx}u_{x} + \frac{1}{2}a_{1}\mathbb{X}_{2}u_{x}^{2} - \frac{1}{2}[\mathbb{X}_{2},(a_{6}\mathbb{X}_{2})_{x}]u_{x}^{2} - \frac{1}{2}a_{2}\mathbb{X}_{2}u_{x}^{2} + (a_{6}[\mathbb{X}_{2},[\mathbb{X}_{1},\mathbb{X}_{2}]])_{x}uu_{x} \nonumber \\
&& - (a_{6}[\mathbb{X}_{1},\mathbb{X}_{2}])_{xx}u_{x} - ([\mathbb{X}_{1},(a_{6}\mathbb{X}_{2})_{x}])_{x}u_{x} - ([\mathbb{X}_{2},(a_{6}\mathbb{X}_{2})_{x}])_{x}uu_{x} + (a_{6}[\mathbb{X}_{1},[\mathbb{X}_{1},\mathbb{X}_{2}]])_{x}u_{x} \nonumber \\
&& - a_{1}[\mathbb{X}_{1},\mathbb{X}_{2}]uu_{x} + \frac{1}{2}a_{6}[\mathbb{X}_{2},[\mathbb{X}_{1},\mathbb{X}_{2}]]u_{x}^{2} + [\mathbb{X}_{2},(a_{6}[\mathbb{X}_{1},\mathbb{X}_{2}])_{x}]uu_{x} + (a_{5}\mathbb{X}_{2})_{x}u_{x} \nonumber \\
&& - a_{5}[\mathbb{X}_{1},\mathbb{X}_{2}]u_{x} - [\mathbb{X}_{1},(a_{6}\mathbb{X}_{2})_{xx}]u_{x} - [\mathbb{X}_{2},(a_{6}\mathbb{X}_{2})_{xx}]uu_{x} + [\mathbb{X}_{1},(a_{6}[\mathbb{X}_{1},\mathbb{X}_{2}])_{x}]u_{x} \nonumber \\
&& + [\mathbb{X}_{1},[\mathbb{X}_{1},(a_{6}\mathbb{X}_{2})_{x}]]u_{x} + [\mathbb{X}_{1},[\mathbb{X}_{2},(a_{6}\mathbb{X}_{2})_{x}]]uu_{x} + [\mathbb{X}_{2},[\mathbb{X}_{1},(a_{6}\mathbb{X}_{2})_{x}]]uu_{x} + (a_{1}\mathbb{X}_{2})_{x}uu_{x} \nonumber \\
&& - a_{6}[\mathbb{X}_{1},[\mathbb{X}_{1},[\mathbb{X}_{1},\mathbb{X}_{2}]]]u_{x} - a_{6}[\mathbb{X}_{1},[\mathbb{X}_{2},[\mathbb{X}_{1},\mathbb{X}_{2}]]]uu_{x} - a_{6}[\mathbb{X}_{2},[\mathbb{X}_{1},[\mathbb{X}_{1},\mathbb{X}_{2}]]]uu_{x} \nonumber \\
&& + [\mathbb{X}_{2},[\mathbb{X}_{2},(a_{6}\mathbb{X}_{2})_{x}]]u^{2}u_{x} - a_{6}[\mathbb{X}_{2},[\mathbb{X}_{2},[\mathbb{X}_{1},\mathbb{X}_{2}]]]u^{2}u_{x} + \mathbb{K}^{3}(x,t,u) \label{KDV8}
\end{eqnarray}

It is helpful at this stage to define the following new matrices

\begin{eqnarray}
&& \mathbb{X}_{4} = [\mathbb{X}_{1},\mathbb{X}_{2}], \ \ \mathbb{X}_{5} = [\mathbb{X}_{1},\mathbb{X}_{4}], \ \ \mathbb{X}_{6} = [\mathbb{X}_{2},\mathbb{X}_{4}] \\
&& \mathbb{X}_{7} = [\mathbb{X}_{1},\mathbb{X}_{5}], \ \ \mathbb{X}_{8} = [\mathbb{X}_{2},\mathbb{X}_{5}], \ \ \mathbb{X}_{9} = [\mathbb{X}_{1},\mathbb{X}_{6}], \ \ \mathbb{X}_{10} = [\mathbb{X}_{2},\mathbb{X}_{6}]
\end{eqnarray}

Now we update $\eqref{KDV6}$ by plugging in $\eqref{KDV8}$. This yields a long expression which is \eqref{KDV9} in Appendix A

Now since $\mathbb{K}^{3}$ and the $\mathbb{X}_{i}$ do not depend on $u_{x}$ we can set the coefficient of the $u_{x}^{2}$ term to zero \eqref{KDV9}. Therefore we require

\begin{eqnarray}
&& - \frac{1}{2}(a_{1}\mathbb{X}_{2})_{x} + \frac{1}{2}([\mathbb{X}_{2},(a_{6}\mathbb{X}_{2})_{x}])_{x} + \frac{1}{2}(a_{2}\mathbb{X}_{2})_{x} - (a_{6}\mathbb{X}_{6})_{x} + a_{1}\mathbb{X}_{4} \nonumber \\
&& - \frac{1}{2}(a_{6}\mathbb{X}_{6})_{x} + ([\mathbb{X}_{2},(a_{6}\mathbb{X}_{2})_{x}])_{x} + a_{6}\mathbb{X}_{9} + [\mathbb{X}_{2},(a_{6}\mathbb{X}_{2})_{xx}] + \frac{1}{2}a_{6}\mathbb{X}_{10}u \nonumber \\
&& - [\mathbb{X}_{1},[\mathbb{X}_{2},(a_{6}\mathbb{X}_{2})_{x}]] - [\mathbb{X}_{2},[\mathbb{X}_{1},(a_{6}\mathbb{X}_{2})_{x}]] - (a_{1}\mathbb{X}_{2})_{x} - [\mathbb{X}_{2},(a_{6}\mathbb{X}_{4})_{x}] \nonumber \\
&& - 2[\mathbb{X}_{2},[\mathbb{X}_{2},(a_{6}\mathbb{X}_{2})_{x}]]u + 2a_{6}\mathbb{X}_{10}u - \frac{1}{2}a_{2}\mathbb{X}_{4} + a_{6}\mathbb{X}_{8} + \frac{1}{2}a_{1}\mathbb{X}_{4} \nonumber \\
&& - \frac{1}{2}[\mathbb{X}_{1},[\mathbb{X}_{2},(a_{6}\mathbb{X}_{2})_{x}]] + \frac{1}{2}a_{6}\mathbb{X}_{9} - \frac{1}{2}[\mathbb{X}_{2},[\mathbb{X}_{2},(a_{6}\mathbb{X}_{2})_{x}]]u = 0 \label{KDV10}
\end{eqnarray}

Further since we know that the $\mathbb{X}_{i}$ do not depend on $u$ we can decouple this condition as follows.

\begin{eqnarray}
&& \frac{3}{2}([\mathbb{X}_{2},(a_{6}\mathbb{X}_{2})_{x}])_{x} - \frac{3}{2}(a_{1}\mathbb{X}_{2})_{x} + \frac{1}{2}(a_{2}\mathbb{X}_{2})_{x} - \frac{3}{2}(a_{6}\mathbb{X}_{6})_{x} + \frac{3}{2}a_{1}\mathbb{X}_{4} + \frac{3}{2}a_{6}\mathbb{X}_{9} \nonumber \\
&& - \frac{3}{2}[\mathbb{X}_{1},[\mathbb{X}_{2},(a_{6}\mathbb{X}_{2})_{x}]] - [\mathbb{X}_{2},[\mathbb{X}_{1},(a_{6}\mathbb{X}_{2})_{x}]] - [\mathbb{X}_{2},(a_{6}\mathbb{X}_{4})_{x}] + [\mathbb{X}_{2},(a_{6}\mathbb{X}_{2})_{xx}] \nonumber \\
&& - \frac{1}{2}a_{2}\mathbb{X}_{4} + a_{6}\mathbb{X}_{8} = 0  \label{KDV11} \\
&& a_{6}\mathbb{X}_{10} - [\mathbb{X}_{2},[\mathbb{X}_{2},(a_{6}\mathbb{X}_{2})_{x}]] = 0 \label{KDV12}
\end{eqnarray}

Taking these conditions into account and once again noting the fact that $\mathbb{K}^{3}$ and the $\mathbb{X}_{i}$ are not independent of $u_{x}$ we can simplify and equate the coefficient of the $u_{x}$ in $\eqref{KDV9}$ to zero. Thus we now obtain the condition \eqref{KDV13} in Appendix A.

Now we update $\eqref{KDV9}$ by plugging in $\eqref{KDV14}$. Upon doing this we will have a rather large expression in which is no more than a algebraic equation in $u$. We will find our remaining constraints by equating the coefficients of the different powers of $u$ in this expression to zero. This updated version of $\eqref{KDV9}$ is very lengthy, and omitted here.

Now, in the final step, as the $\mathbb{X}_{i}$ do not depend on $u$ we can set the coefficients of the different powers of $u$ in this last, lengthy expression to zero. Thus we have

\begin{eqnarray}
O(1) &:& \mathbb{X}_{1,t} - \mathbb{X}_{0,x} + [\mathbb{X}_{1},\mathbb{X}_{0}] = 0 \\
O(u) &:& [\mathbb{X}_{2},\mathbb{X}_{0}] - a_{6}[\mathbb{X}_{1},[\mathbb{X}_{1},\mathbb{X}_{7}]] + [\mathbb{X}_{1},[\mathbb{X}_{1},[\mathbb{X}_{1},(a_{6}\mathbb{X}_{4})_{x}]]] + [\mathbb{X}_{1},[\mathbb{X}_{1},[\mathbb{X}_{1},[\mathbb{X}_{1},(a_{6}\mathbb{X}_{2})_{x}]]]] \nonumber \\
&& + [\mathbb{X}_{1},[\mathbb{X}_{1},(a_{5}\mathbb{X}_{2})_{x}]] - [\mathbb{X}_{1},[\mathbb{X}_{1},[\mathbb{X}_{1},(a_{6}\mathbb{X}_{2})_{xx}]]] + [\mathbb{X}_{1},(a_{6}\mathbb{X}_{7})_{x}] - [\mathbb{X}_{1},[\mathbb{X}_{1},(a_{6}\mathbb{X}_{4})_{xx}]] \nonumber \\
&& - a_{5}\mathbb{X}_{7} - [\mathbb{X}_{1},[\mathbb{X}_{1},([\mathbb{X}_{1},(a_{6}\mathbb{X}_{2})_{x}])_{x}]] + [\mathbb{X}_{1},[\mathbb{X}_{1},(a_{6}\mathbb{X}_{2})_{xxx}]] - [\mathbb{X}_{1},([\mathbb{X}_{1},(a_{6}\mathbb{X}_{4})_{x}])_{x}] \nonumber \\
&& - [\mathbb{X}_{1},([\mathbb{X}_{1},[\mathbb{X}_{1},(a_{6}\mathbb{X}_{2})_{x}]])_{x}] + [\mathbb{X}_{1},(a_{5}\mathbb{X}_{4})_{x}] + [\mathbb{X}_{1},([\mathbb{X}_{1},(a_{6}\mathbb{X}_{2})_{xx}])_{x}] - [\mathbb{X}_{1},(a_{5}\mathbb{X}_{2})_{xx}] \nonumber \\
&& - [\mathbb{X}_{1},(a_{6}\mathbb{X}_{5})_{xx}] + [\mathbb{X}_{1},(a_{6}\mathbb{X}_{4})_{xxx}] + [\mathbb{X}_{1},([\mathbb{X}_{1},(a_{6}\mathbb{X}_{2})_{x}])_{xx}] - a_{8}\mathbb{X}_{4} - [\mathbb{X}_{1},(a_{6}\mathbb{X}_{2})_{xxxx}] \nonumber \\
&& + [\mathbb{X}_{1},[\mathbb{X}_{1},(a_{6}\mathbb{X}_{5})_{x}]] + (a_{6}[\mathbb{X}_{1},\mathbb{X}_{7}])_{x} - ([\mathbb{X}_{1},[\mathbb{X}_{1},[\mathbb{X}_{1},(a_{6}\mathbb{X}_{2})_{x}]]])_{x} - ([\mathbb{X}_{1},[\mathbb{X}_{1},(a_{6}\mathbb{X}_{4})_{x}]])_{x} \nonumber \\
&& + (a_{8}\mathbb{X}_{2})_{x} + ([\mathbb{X}_{1},[\mathbb{X}_{1},(a_{6}\mathbb{X}_{2})_{xx}]])_{x} - ([\mathbb{X}_{1},(a_{5}\mathbb{X}_{2})_{x}])_{x} - (a_{6}\mathbb{X}_{7})_{xx} + ([\mathbb{X}_{1},(a_{6}\mathbb{X}_{4})_{xx}])_{x} \nonumber \\
&& + ([\mathbb{X}_{1},([\mathbb{X}_{1},(a_{6}\mathbb{X}_{2})_{x}])_{x}])_{x} - ([\mathbb{X}_{1},(a_{6}\mathbb{X}_{2})_{xxx}])_{x} + ([\mathbb{X}_{1},[\mathbb{X}_{1},(a_{6}\mathbb{X}_{2})_{x}]])_{xx} + (a_{5}\mathbb{X}_{5})_{x} \nonumber \\
&& + \mathbb{X}_{2,t} - ([\mathbb{X}_{1},(a_{6}\mathbb{X}_{2})_{xx}])_{xx} - (a_{5}\mathbb{X}_{4})_{xx} + (a_{5}\mathbb{X}_{2})_{xxx} + (a_{6}\mathbb{X}_{5})_{xxx} - ([\mathbb{X}_{1},(a_{6}\mathbb{X}_{2})_{x}])_{xxx} \nonumber \\
&& - ([\mathbb{X}_{1},(a_{6}\mathbb{X}_{5})_{x}])_{x} + (a_{6}\mathbb{X}_{2})_{xxxxx} - (a_{6}\mathbb{X}_{4})_{xxxx} - a_{7}\mathbb{X}_{2} + ([\mathbb{X}_{1},(a_{6}\mathbb{X}_{4})_{x}])_{xx} = 0
\end{eqnarray}

\begin{eqnarray}
O(u^{2}) &:& - a_{5}\mathbb{X}_{8} - a_{6}[\mathbb{X}_{2},[\mathbb{X}_{1},\mathbb{X}_{7}]] + [\mathbb{X}_{2},[\mathbb{X}_{1},[\mathbb{X}_{1},[\mathbb{X}_{1},(a_{6}\mathbb{X}_{2})_{x}]]]] + [\mathbb{X}_{2},[\mathbb{X}_{1},[\mathbb{X}_{1},(a_{6}\mathbb{X}_{4})_{x}]]] \nonumber \\
&& - [\mathbb{X}_{2},[\mathbb{X}_{1},[\mathbb{X}_{1},(a_{6}\mathbb{X}_{2})_{xx}]]] + [\mathbb{X}_{2},[\mathbb{X}_{1},(a_{5}\mathbb{X}_{2})_{x}]] - [\mathbb{X}_{2},[\mathbb{X}_{1},([\mathbb{X}_{1},(a_{6}\mathbb{X}_{2})_{x}])_{x}]] - \frac{1}{2}a_{4}\mathbb{X}_{4} \nonumber \\
&& + [\mathbb{X}_{2},(a_{6}\mathbb{X}_{7})_{x}] - [\mathbb{X}_{2},[\mathbb{X}_{1},(a_{6}\mathbb{X}_{4})_{xx}]] + [\mathbb{X}_{2},[\mathbb{X}_{1},(a_{6}\mathbb{X}_{2})_{xxx}]] - [\mathbb{X}_{2},([\mathbb{X}_{1},(a_{6}\mathbb{X}_{4})_{x}])_{x}] \nonumber \\
&& - [\mathbb{X}_{2},([\mathbb{X}_{1},[\mathbb{X}_{1},(a_{6}\mathbb{X}_{2})_{x}]])_{x}] + [\mathbb{X}_{2},(a_{5}\mathbb{X}_{4})_{x}] + [\mathbb{X}_{2},([\mathbb{X}_{1},(a_{6}\mathbb{X}_{2})_{xx}])_{x}] - [\mathbb{X}_{2},(a_{5}\mathbb{X}_{2})_{xx}] \nonumber \\
&& + [\mathbb{X}_{2},(a_{6}\mathbb{X}_{4})_{xxx}] + [\mathbb{X}_{2},([\mathbb{X}_{1},(a_{6}\mathbb{X}_{2})_{x}])_{xx}] - [\mathbb{X}_{2},(a_{6}\mathbb{X}_{2})_{xxxx}] + [\mathbb{X}_{2},[\mathbb{X}_{1},(a_{6}\mathbb{X}_{5})_{x}]] \nonumber \\
&& - [\mathbb{X}_{2},(a_{6}\mathbb{X}_{5})_{xx}] - \frac{1}{2}a_{5}\mathbb{X}_{9} + \frac{1}{2}[\mathbb{X}_{1},[\mathbb{X}_{2},(a_{5}\mathbb{X}_{2})_{x}]] + \frac{1}{2}[\mathbb{X}_{1},[\mathbb{X}_{2},[\mathbb{X}_{1},[\mathbb{X}_{1},(a_{6}\mathbb{X}_{2})_{x}]]]] \nonumber \\
&& - \frac{1}{2}[\mathbb{X}_{1},[\mathbb{X}_{2},[\mathbb{X}_{1},(a_{6}\mathbb{X}_{2})_{xx}]]] + \frac{1}{2}[\mathbb{X}_{1},[\mathbb{X}_{1},(a_{1}\mathbb{X}_{2})_{x}]] - \frac{1}{2}[\mathbb{X}_{1},[\mathbb{X}_{2},([\mathbb{X}_{1},(a_{6}\mathbb{X}_{2})_{x}])_{x}]] \nonumber \\
&& + \frac{1}{2}[\mathbb{X}_{1},[\mathbb{X}_{1},[\mathbb{X}_{1},[\mathbb{X}_{2},(a_{6}\mathbb{X}_{2})_{x}]]]] - \frac{1}{2}[\mathbb{X}_{1},[\mathbb{X}_{2},(a_{6}\mathbb{X}_{4})_{xx}]] + \frac{1}{2}[\mathbb{X}_{1},[\mathbb{X}_{2},[\mathbb{X}_{1},(a_{6}\mathbb{X}_{4})_{x}]]] \nonumber \\
&& - \frac{1}{2}a_{6}[\mathbb{X}_{1},[\mathbb{X}_{1},\mathbb{X}_{9}]] + \frac{1}{2}[\mathbb{X}_{1},[\mathbb{X}_{2},(a_{6}\mathbb{X}_{2})_{xxx}]] + \frac{1}{2}[\mathbb{X}_{1},[\mathbb{X}_{1},[\mathbb{X}_{2},[\mathbb{X}_{1},(a_{6}\mathbb{X}_{2})_{x}]]]] \nonumber \\
&& + \frac{1}{2}[\mathbb{X}_{1},[\mathbb{X}_{2},(a_{6}\mathbb{X}_{5})_{x}]] - \frac{1}{2}[\mathbb{X}_{1},[\mathbb{X}_{1},[\mathbb{X}_{2},(a_{6}\mathbb{X}_{2})_{xx}]]] - \frac{1}{2}[\mathbb{X}_{1},[\mathbb{X}_{1},([\mathbb{X}_{2},(a_{6}\mathbb{X}_{2})_{x}])_{x}]] \nonumber \\
&& - \frac{1}{2}a_{6}[\mathbb{X}_{1},[\mathbb{X}_{1},\mathbb{X}_{8}]] + \frac{1}{2}[\mathbb{X}_{1},[\mathbb{X}_{1},[\mathbb{X}_{2},(a_{6}\mathbb{X}_{4})_{x}]]] - \frac{1}{2}a_{1}\mathbb{X}_{7} + \frac{1}{2}[\mathbb{X}_{1},[\mathbb{X}_{1},(a_{6}\mathbb{X}_{6})_{x}]] \nonumber \\
&& - \frac{1}{2}[\mathbb{X}_{1},([\mathbb{X}_{2},[\mathbb{X}_{1},(a_{6}\mathbb{X}_{2})_{x}]])_{x}] + \frac{1}{2}[\mathbb{X}_{1},(a_{6}\mathbb{X}_{9})_{x}] + \frac{1}{2}[\mathbb{X}_{1},(a_{6}\mathbb{X}_{8})_{x}] - \frac{1}{2}[\mathbb{X}_{1},(a_{6}\mathbb{X}_{6})_{xx}] \nonumber \\
&& - \frac{1}{2}[\mathbb{X}_{1},([\mathbb{X}_{1},[\mathbb{X}_{2},(a_{6}\mathbb{X}_{2})_{x}]])_{x}] + \frac{1}{2}[\mathbb{X}_{1},([\mathbb{X}_{2},(a_{6}\mathbb{X}_{2})_{xx}])_{x}] - \frac{1}{2}[\mathbb{X}_{1},([\mathbb{X}_{2},(a_{6}\mathbb{X}_{4})_{x}])_{x}] \nonumber \\
&& - \frac{1}{2}[\mathbb{X}_{1},(a_{1}\mathbb{X}_{2})_{xx}] + \frac{1}{2}[\mathbb{X}_{1},(a_{1}\mathbb{X}_{4})_{x}] + \frac{1}{2}[\mathbb{X}_{1},([\mathbb{X}_{2},(a_{6}\mathbb{X}_{2})_{x}])_{xx}] - \frac{1}{2}a_{6}[\mathbb{X}_{1},[\mathbb{X}_{2},\mathbb{X}_{7}]] \nonumber \\
&& - \frac{1}{2}([\mathbb{X}_{2},(a_{5}\mathbb{X}_{2})_{x}])_{x} - \frac{1}{2}([\mathbb{X}_{2},[\mathbb{X}_{1},[\mathbb{X}_{1},(a_{6}\mathbb{X}_{2})_{x}]]])_{x} + \frac{1}{2}([\mathbb{X}_{2},[\mathbb{X}_{1},(a_{6}\mathbb{X}_{2})_{xx}]])_{x} \nonumber \\
&& + \frac{1}{2}(a_{5}\mathbb{X}_{6})_{x} + \frac{1}{2}([\mathbb{X}_{2},([\mathbb{X}_{1},(a_{6}\mathbb{X}_{2})_{x}])_{x}])_{x} - \frac{1}{2}([\mathbb{X}_{2},(a_{6}\mathbb{X}_{5})_{x}])_{x} - \frac{1}{2}([\mathbb{X}_{1},(a_{1}\mathbb{X}_{2})_{x}])_{x} \nonumber \\
&& + \frac{1}{2}([\mathbb{X}_{2},(a_{6}\mathbb{X}_{4})_{xx}])_{x} + \frac{1}{2}(a_{6}[\mathbb{X}_{1},\mathbb{X}_{9}])_{x} + \frac{1}{2}(a_{6}[\mathbb{X}_{1},\mathbb{X}_{8}])_{x} - \frac{1}{2}([\mathbb{X}_{2},[\mathbb{X}_{1},(a_{6}\mathbb{X}_{4})_{x}]])_{x} \nonumber \\
&& - \frac{1}{2}([\mathbb{X}_{2},(a_{6}\mathbb{X}_{2})_{xxx}])_{x} - \frac{1}{2}([\mathbb{X}_{1},[\mathbb{X}_{1},[\mathbb{X}_{2},(a_{6}\mathbb{X}_{2})_{x}]]])_{x} - \frac{1}{2}([\mathbb{X}_{1},[\mathbb{X}_{2},[\mathbb{X}_{1},(a_{6}\mathbb{X}_{2})_{x}]]])_{x} \nonumber \\
&& + \frac{1}{2}([\mathbb{X}_{1},[\mathbb{X}_{2},(a_{6}\mathbb{X}_{2})_{xx}]])_{x} - \frac{1}{2}([\mathbb{X}_{1},[\mathbb{X}_{2},(a_{6}\mathbb{X}_{4})_{x}]])_{x} + \frac{1}{2}([\mathbb{X}_{1},([\mathbb{X}_{2},(a_{6}\mathbb{X}_{2})_{x}])_{x}])_{x} \nonumber \\
&& + \frac{1}{2}(a_{1}\mathbb{X}_{5})_{x} - \frac{1}{2}([\mathbb{X}_{1},(a_{6}\mathbb{X}_{6})_{x}])_{x} - \frac{1}{2}(a_{6}\mathbb{X}_{9})_{xx} - \frac{1}{2}(a_{6}\mathbb{X}_{8})_{xx} + \frac{1}{2}([\mathbb{X}_{1},[\mathbb{X}_{2},(a_{6}\mathbb{X}_{2})_{x}]])_{xx} \nonumber \\
&& + \frac{1}{2}(a_{6}\mathbb{X}_{6})_{xxx} + \frac{1}{2}([\mathbb{X}_{2},[\mathbb{X}_{1},(a_{6}\mathbb{X}_{2})_{x}]])_{xx} - \frac{1}{2}([\mathbb{X}_{2},(a_{6}\mathbb{X}_{2})_{xx}])_{xx} + \frac{1}{2}([\mathbb{X}_{2},(a_{6}\mathbb{X}_{4})_{x}])_{xx} \nonumber \\
&& + \frac{1}{2}(a_{1}\mathbb{X}_{2})_{xxx} - \frac{1}{2}(a_{1}\mathbb{X}_{4})_{xx} - \frac{1}{2}([\mathbb{X}_{2},(a_{6}\mathbb{X}_{2})_{x}])_{xxx} + \frac{1}{2}(a_{6}[\mathbb{X}_{2},\mathbb{X}_{7}])_{x} + \frac{1}{2}(a_{4}\mathbb{X}_{2})_{x} = 0
\end{eqnarray}

\begin{eqnarray}
O(u^{3}) &:& - \frac{1}{3}([\mathbb{X}_{2},(a_{6}\mathbb{X}_{6})_{x}])_{x} + \frac{1}{2}[\mathbb{X}_{2},[\mathbb{X}_{2},[\mathbb{X}_{1},[\mathbb{X}_{1},(a_{6}\mathbb{X}_{2})_{x}]]]] - \frac{1}{2}[\mathbb{X}_{2},[\mathbb{X}_{2},[\mathbb{X}_{1},(a_{6}\mathbb{X}_{2})_{xx}]]] \nonumber \\
&& + \frac{1}{3}(a_{3}\mathbb{X}_{2})_{x} - \frac{1}{2}[\mathbb{X}_{2},[\mathbb{X}_{2},(a_{6}\mathbb{X}_{4})_{xx}]] + \frac{1}{2}[\mathbb{X}_{2},[\mathbb{X}_{2},(a_{6}\mathbb{X}_{5})_{x}]] + \frac{1}{2}[\mathbb{X}_{2},[\mathbb{X}_{1},(a_{1}\mathbb{X}_{2})_{x}]] \nonumber \\
&& - \frac{1}{2}[\mathbb{X}_{2},[\mathbb{X}_{2},([\mathbb{X}_{1},(a_{6}\mathbb{X}_{2})_{x}])_{x}]] - \frac{1}{2}a_{6}[\mathbb{X}_{2},[\mathbb{X}_{1},\mathbb{X}_{9}]] + \frac{1}{2}[\mathbb{X}_{2},[\mathbb{X}_{1},[\mathbb{X}_{2},[\mathbb{X}_{1},(a_{6}\mathbb{X}_{2})_{x}]]]] \nonumber \\
&& - \frac{1}{2}a_{6}[\mathbb{X}_{2},[\mathbb{X}_{1},\mathbb{X}_{8}]] + \frac{1}{2}[\mathbb{X}_{2},[\mathbb{X}_{2},[\mathbb{X}_{1},(a_{6}\mathbb{X}_{4})_{x}]]] + \frac{1}{2}[\mathbb{X}_{2},[\mathbb{X}_{1},[\mathbb{X}_{1},[\mathbb{X}_{2},(a_{6}\mathbb{X}_{2})_{x}]]]] \nonumber \\
&& + \frac{1}{2}[\mathbb{X}_{2},[\mathbb{X}_{2},(a_{6}\mathbb{X}_{2})_{xxx}]] - \frac{1}{2}[\mathbb{X}_{2},[\mathbb{X}_{1},[\mathbb{X}_{2},(a_{6}\mathbb{X}_{2})_{xx}]]] - \frac{1}{2}[\mathbb{X}_{2},[\mathbb{X}_{1},([\mathbb{X}_{2},(a_{6}\mathbb{X}_{2})_{x}])_{x}]] \nonumber \\
&& - \frac{1}{2}a_{1}\mathbb{X}_{8} + \frac{1}{2}[\mathbb{X}_{2},[\mathbb{X}_{1},[\mathbb{X}_{2},(a_{6}\mathbb{X}_{4})_{x}]]] + \frac{1}{2}[\mathbb{X}_{2},[\mathbb{X}_{1},(a_{6}\mathbb{X}_{6})_{x}]] - \frac{1}{2}[\mathbb{X}_{2},(a_{6}\mathbb{X}_{6})_{xx}] \nonumber \\
&& - \frac{1}{2}[\mathbb{X}_{2},([\mathbb{X}_{2},[\mathbb{X}_{1},(a_{6}\mathbb{X}_{2})_{x}]])_{x}] + \frac{1}{2}[\mathbb{X}_{2},(a_{6}\mathbb{X}_{9})_{x}] - \frac{1}{2}[\mathbb{X}_{2},([\mathbb{X}_{1},[\mathbb{X}_{2},(a_{6}\mathbb{X}_{2})_{x}]])_{x}] \nonumber \\
&& + \frac{1}{2}[\mathbb{X}_{2},([\mathbb{X}_{2},(a_{6}\mathbb{X}_{2})_{xx}])_{x}] - \frac{1}{2}[\mathbb{X}_{2},(a_{1}\mathbb{X}_{2})_{xx}] + \frac{1}{2}[\mathbb{X}_{2},([\mathbb{X}_{2},(a_{6}\mathbb{X}_{2})_{x}])_{xx}] + \frac{1}{2}[\mathbb{X}_{2},(a_{1}\mathbb{X}_{4})_{x}] \nonumber \\
&& + \frac{1}{2}[\mathbb{X}_{2},(a_{6}\mathbb{X}_{8})_{x}] - \frac{1}{2}[\mathbb{X}_{2},([\mathbb{X}_{2},(a_{6}\mathbb{X}_{4})_{x}])_{x}] - \frac{1}{2}a_{6}[\mathbb{X}_{2},[\mathbb{X}_{2},\mathbb{X}_{7}]] - \frac{1}{3}a_{6}[\mathbb{X}_{1},[\mathbb{X}_{2},\mathbb{X}_{9}]] \nonumber \\
&& + \frac{1}{3}[\mathbb{X}_{1},[\mathbb{X}_{2},(a_{1}\mathbb{X}_{2})_{x}]] + \frac{1}{3}[\mathbb{X}_{1},[\mathbb{X}_{2},[\mathbb{X}_{2},[\mathbb{X}_{1},(a_{6}\mathbb{X}_{2})_{x}]]]] - \frac{1}{3}[\mathbb{X}_{1},[\mathbb{X}_{2},([\mathbb{X}_{2},(a_{6}\mathbb{X}_{2})_{x}])_{x}]] \nonumber \\
&& + \frac{1}{3}[\mathbb{X}_{1},[\mathbb{X}_{2},[\mathbb{X}_{1},[\mathbb{X}_{2},(a_{6}\mathbb{X}_{2})_{x}]]]] - \frac{1}{3}[\mathbb{X}_{1},[\mathbb{X}_{2},[\mathbb{X}_{2},(a_{6}\mathbb{X}_{2})_{xx}]]] - \frac{1}{3}a_{6}[\mathbb{X}_{1},[\mathbb{X}_{2},\mathbb{X}_{8}]] \nonumber \\
&& + \frac{1}{3}[\mathbb{X}_{1},[\mathbb{X}_{2},(a_{6}\mathbb{X}_{6})_{x}]] + \frac{1}{3}[\mathbb{X}_{1},[\mathbb{X}_{2},[\mathbb{X}_{2},(a_{6}\mathbb{X}_{4})_{x}]]] - \frac{1}{3}a_{1}\mathbb{X}_{9} - \frac{1}{3}([\mathbb{X}_{2},(a_{1}\mathbb{X}_{2})_{x}])_{x} \nonumber \\
&& - \frac{1}{3}a_{3}\mathbb{X}_{4} + \frac{1}{3}(a_{6}[\mathbb{X}_{2},\mathbb{X}_{9}])_{x} + \frac{1}{3}(a_{6}[\mathbb{X}_{2},\mathbb{X}_{8}])_{x} - \frac{1}{3}([\mathbb{X}_{2},[\mathbb{X}_{1},[\mathbb{X}_{2},(a_{6}\mathbb{X}_{2})_{x}]]])_{x} \nonumber \\
&& - \frac{1}{3}([\mathbb{X}_{2},[\mathbb{X}_{2},[\mathbb{X}_{1},(a_{6}\mathbb{X}_{2})_{x}]]])_{x} + \frac{1}{3}([\mathbb{X}_{2},[\mathbb{X}_{2},(a_{6}\mathbb{X}_{2})_{xx}]])_{x} + \frac{1}{3}([\mathbb{X}_{2},([\mathbb{X}_{2},(a_{6}\mathbb{X}_{2})_{x}])_{x}])_{x} \nonumber \\
&& - \frac{1}{3}([\mathbb{X}_{2},[\mathbb{X}_{2},(a_{6}\mathbb{X}_{4})_{x}]])_{x} + \frac{1}{3}(a_{1}\mathbb{X}_{6})_{x} = 0
\end{eqnarray}

\begin{eqnarray}
O(u^{4}) &:& [\mathbb{X}_{2},[\mathbb{X}_{2},[\mathbb{X}_{2},[\mathbb{X}_{1},(a_{6}\mathbb{X}_{2})_{x}]]]] - a_{6}[\mathbb{X}_{2},[\mathbb{X}_{2},\mathbb{X}_{8}]] + [\mathbb{X}_{2},[\mathbb{X}_{2},[\mathbb{X}_{1},[\mathbb{X}_{2},(a_{6}\mathbb{X}_{2})_{x}]]]] \nonumber \\
&& - a_{6}[\mathbb{X}_{2},[\mathbb{X}_{2},\mathbb{X}_{9}]] - [\mathbb{X}_{2},[\mathbb{X}_{2},[\mathbb{X}_{2},(a_{6}\mathbb{X}_{2})_{xx}]]] - [\mathbb{X}_{2},[\mathbb{X}_{2},([\mathbb{X}_{2},(a_{6}\mathbb{X}_{2})_{x}])_{x}]] \nonumber \\
&& + [\mathbb{X}_{2},[\mathbb{X}_{2},(a_{6}\mathbb{X}_{6})_{x}]] + [\mathbb{X}_{2},[\mathbb{X}_{2},[\mathbb{X}_{2},(a_{6}\mathbb{X}_{4})_{x}]]] = 0
\end{eqnarray}

   Note that if we decouple $\eqref{KDV11}$ into the following conditions

\begin{eqnarray}
&& ([\mathbb{X}_{2},(a_{6}\mathbb{X}_{2})_{x}])_{x} - (a_{6}\mathbb{X}_{6})_{x} + a_{6}\mathbb{X}_{9} - [\mathbb{X}_{1},[\mathbb{X}_{2},(a_{6}\mathbb{X}_{2})_{x}]] = 0 \label{NEWKDVCON1} \\
&& [\mathbb{X}_{2},[\mathbb{X}_{1},(a_{6}\mathbb{X}_{2})_{x}]] + [\mathbb{X}_{2},(a_{6}\mathbb{X}_{4})_{x}] - [\mathbb{X}_{2},(a_{6}\mathbb{X}_{2})_{xx}] - a_{6}\mathbb{X}_{8} = 0 \label{NEWKDVCON2} \\
&& ((a_{2} - 3a_{1})\mathbb{X}_{2})_{x} - (a_{2} - 3a_{1})\mathbb{X}_{4} = 0 \label{NEWKDVCON3}
\end{eqnarray}

\noindent
then the $O(u^{4})$ equation is identically satisfied. To reduce the complexity of the $O(u^{3})$ equation we can decouple it into the following equations

\begin{eqnarray}
&& [\mathbb{X}_{2},[\mathbb{X}_{1},[\mathbb{X}_{1},(a_{6}\mathbb{X}_{2})_{x}]]] - [\mathbb{X}_{2},[\mathbb{X}_{1},(a_{6}\mathbb{X}_{2})_{xx}]] - [\mathbb{X}_{2},(a_{6}\mathbb{X}_{4})_{xx}] + [\mathbb{X}_{2},(a_{6}\mathbb{X}_{5})_{x}] \nonumber \\
&& + [\mathbb{X}_{1},(a_{1}\mathbb{X}_{2})_{x}] - [\mathbb{X}_{2},([\mathbb{X}_{1},(a_{6}\mathbb{X}_{2})_{x}])_{x}] + [\mathbb{X}_{2},[\mathbb{X}_{1},(a_{6}\mathbb{X}_{4})_{x}]] + [\mathbb{X}_{2},(a_{6}\mathbb{X}_{2})_{xxx}] \nonumber \\
&& - a_{1}\mathbb{X}_{5} - (a_{1}\mathbb{X}_{2})_{xx} + (a_{1}\mathbb{X}_{4})_{x} - a_{6}[\mathbb{X}_{2},\mathbb{X}_{7}] = 0 \label{NEWKDVCON4} \\
&& (a_{3}\mathbb{X}_{2})_{x} + [\mathbb{X}_{1},[\mathbb{X}_{2},(a_{1}\mathbb{X}_{2})_{x}]] - a_{1}\mathbb{X}_{9} - ([\mathbb{X}_{2},(a_{1}\mathbb{X}_{2})_{x}])_{x} - a_{3}\mathbb{X}_{4} + (a_{1}\mathbb{X}_{6})_{x} = 0 \label{NEWKDVCON5}
\end{eqnarray}

\noindent
From this last condition, we can use $\eqref{NEWKDVCON1}-\eqref{NEWKDVCON5}$ to reduce the $O(u^{2})$ condition to the following

\begin{eqnarray}
&& - a_{5}\mathbb{X}_{8} - a_{6}[\mathbb{X}_{2},[\mathbb{X}_{1},\mathbb{X}_{7}]] + [\mathbb{X}_{2},[\mathbb{X}_{1},[\mathbb{X}_{1},[\mathbb{X}_{1},(a_{6}\mathbb{X}_{2})_{x}]]]] + [\mathbb{X}_{2},[\mathbb{X}_{1},[\mathbb{X}_{1},(a_{6}\mathbb{X}_{4})_{x}]]] \nonumber \\
&& - [\mathbb{X}_{2},[\mathbb{X}_{1},[\mathbb{X}_{1},(a_{6}\mathbb{X}_{2})_{xx}]]] + [\mathbb{X}_{2},[\mathbb{X}_{1},(a_{5}\mathbb{X}_{2})_{x}]] - [\mathbb{X}_{2},[\mathbb{X}_{1},([\mathbb{X}_{1},(a_{6}\mathbb{X}_{2})_{x}])_{x}]] \nonumber \\
&& + [\mathbb{X}_{2},(a_{6}\mathbb{X}_{7})_{x}] - [\mathbb{X}_{2},[\mathbb{X}_{1},(a_{6}\mathbb{X}_{4})_{xx}]] + [\mathbb{X}_{2},[\mathbb{X}_{1},(a_{6}\mathbb{X}_{2})_{xxx}]] - [\mathbb{X}_{2},([\mathbb{X}_{1},(a_{6}\mathbb{X}_{4})_{x}])_{x}] \nonumber \\
&& - [\mathbb{X}_{2},([\mathbb{X}_{1},[\mathbb{X}_{1},(a_{6}\mathbb{X}_{2})_{x}]])_{x}] + [\mathbb{X}_{2},(a_{5}\mathbb{X}_{4})_{x}] + [\mathbb{X}_{2},([\mathbb{X}_{1},(a_{6}\mathbb{X}_{2})_{xx}])_{x}] - [\mathbb{X}_{2},(a_{5}\mathbb{X}_{2})_{xx}] \nonumber \\
&& + [\mathbb{X}_{2},(a_{6}\mathbb{X}_{4})_{xxx}] + [\mathbb{X}_{2},([\mathbb{X}_{1},(a_{6}\mathbb{X}_{2})_{x}])_{xx}] - [\mathbb{X}_{2},(a_{6}\mathbb{X}_{2})_{xxxx}] + [\mathbb{X}_{2},[\mathbb{X}_{1},(a_{6}\mathbb{X}_{5})_{x}]] \nonumber \\
&& - [\mathbb{X}_{2},(a_{6}\mathbb{X}_{5})_{xx}] - \frac{1}{2}a_{5}\mathbb{X}_{9} + \frac{1}{2}[\mathbb{X}_{1},[\mathbb{X}_{2},(a_{5}\mathbb{X}_{2})_{x}]] - \frac{1}{2}([\mathbb{X}_{2},(a_{5}\mathbb{X}_{2})_{x}])_{x} + \frac{1}{2}(a_{5}\mathbb{X}_{6})_{x} \nonumber \\
&& - \frac{1}{2}a_{4}\mathbb{X}_{4} + \frac{1}{2}(a_{4}\mathbb{X}_{2})_{x} = 0
\end{eqnarray}

Decoupling this equation allows for the simplification of the $O(u)$ equation. Thus we write the previous condition as the following system of equations

\begin{eqnarray}
&& - [\mathbb{X}_{1},(a_{6}\mathbb{X}_{4})_{x}] - [\mathbb{X}_{1},[\mathbb{X}_{1},(a_{6}\mathbb{X}_{2})_{x}]] + a_{5}\mathbb{X}_{4} + [\mathbb{X}_{1},(a_{6}\mathbb{X}_{2})_{xx}] - (a_{5}\mathbb{X}_{2})_{x} \nonumber \\
&& + (a_{6}\mathbb{X}_{4})_{xx} + ([\mathbb{X}_{1},(a_{6}\mathbb{X}_{2})_{x}])_{x} - (a_{6}\mathbb{X}_{2})_{xxx} - (a_{6}\mathbb{X}_{5})_{x} = 0 \label{NEWKDVCON6} \\
&& - a_{5}\mathbb{X}_{9} + [\mathbb{X}_{1},[\mathbb{X}_{2},(a_{5}\mathbb{X}_{2})_{x}]] - ([\mathbb{X}_{2},(a_{5}\mathbb{X}_{2})_{x}])_{x} + (a_{5}\mathbb{X}_{6})_{x} - \frac{1}{2}a_{4}\mathbb{X}_{4} + \frac{1}{2}(a_{4}\mathbb{X}_{2})_{x} = 0 \label{NEWKDVCON7}
\end{eqnarray}

Using this the $O(u)$ equation is reduced to

\begin{equation}
\mathbb{X}_{2,t} + [\mathbb{X}_{2},\mathbb{X}_{0}] - a_{8}\mathbb{X}_{4} + (a_{8}\mathbb{X}_{2})_{x} - a_{7}\mathbb{X}_{2} = 0 \label{NEWKDVCON8}
\end{equation}

We therefore find that the final, reduced constraints are given by $\eqref{KDV12}, \eqref{NEWKDVCON1}-\eqref{NEWKDVCON5}$ and $\eqref{NEWKDVCON6}-\eqref{NEWKDVCON8}$. In order to satisfy these constraints we begin with the following rather simple forms for our generators,

\[ \mathbb{X}_{0} = \begin{bmatrix}
g_{1}(x,t) & g_{2}(x,t) \\
g_{3}(x,t) & g_{4}(x,t)
\end{bmatrix}, \ \ \mathbb{X}_{1} = \begin{bmatrix}
0 & f_{1}(x,t) \\
f_{2}(x,t) & 0
\end{bmatrix}, \ \ \mathbb{X}_{2} = \begin{bmatrix}
0 & f_{3}(x,t) \\
f_{4}(x,t) & 0
\end{bmatrix} \]

{\it To get more general results we will assume $a_{2} \neq 3a_{1}$. Note that had we instead opted for the forms}

\[ \mathbb{X}_{0} = \begin{bmatrix}
g_{1}(x,t) & g_{12}(x,t) \\
g_{23}(x,t) & g_{34}(x,t)
\end{bmatrix}, \ \ \mathbb{X}_{1} = \begin{bmatrix}
f_{1}(x,t) & f_{3}(x,t) \\
f_{5}(x,t) & f_{7}(x,t)
\end{bmatrix}, \ \ \mathbb{X}_{2} = \begin{bmatrix}
f_{2}(x,t) & f_{4}(x,t) \\
f_{6}(x,t) & f_{8}(x,t)
\end{bmatrix} \]

\noindent
{\it we would obtain an equivalent system to that obtained in $\cite{Lecce}$. The additional unknown functions which appear in Khawaja's method $\cite{Lecce}$ can be introduced with the proper substitutions via their functional dependence on the twelve unknown functions given above.}

Taking the naive approach of beginning with the smaller conditions first we begin with $\eqref{NEWKDVCON3}$ which, utilizing the given forms for $\mathbb{X}_{0}, \mathbb{X}_{1}$, and $\mathbb{X}_{2}$, becomes

\begin{eqnarray}
&& (a_{2}-3a_{1})(f_{1}f_{4}-f_{2}f_{3}) = 0 \\
&& ((a_{2}-3a_{1})f_{j})_{x} = 0 \ \ \ j=3,4
\end{eqnarray}

\noindent
Solving this system for $f_{2},f_{3}$ and $f_{4}$ yields

\begin{eqnarray}
f_{3}(x,t) &=& \frac{F_{3}(t)}{a_{2}(x,t)-3a_{1}(x,t)} \\
f_{4}(x,t) &=& \frac{F_{4}(t)}{a_{2}(x,t)-3a_{1}(x,t)} \\
f_{2}(x,t) &=& \frac{f_{1}(x,t)F_{4}(t)}{F_{3}(t)} \\
\end{eqnarray}

\noindent
where $F_{3,4}(t)$ are arbitrary functions of $t$. With these choices we've elected to satisfy $\mathbb{X}_{4} = 0$ rather than $a_{2} = 3a_{1}$. Looking next at $\eqref{NEWKDVCON8}$ we obtain the system 

\begin{eqnarray}
&& \left(\frac{F_{j}}{a_{2}-3a_{1}}\right)_{t} - \frac{F_{j}a_{7}}{a_{2}-3a_{1}} + \left(\frac{F_{j}a_{8}}{a_{2}-3a_{1}}\right)_{x} + \frac{1}{2}\left(\frac{F_{j}a_{4}}{a_{2}-3a_{1}}\right)_{x} \nonumber \\
&& + (-1)^{j}\frac{F_{j}(g_{4}-g_{1})}{a_{2}-3a_{1}} = 0 \ \ \ j=3,4 \label{KDVFINAL1}\\
&& \frac{F_{3}g_{3}}{a_{2}-3a_{1}} - \frac{F_{4}g_{2}}{a_{2}-3a_{1}} = 0
\end{eqnarray}

\noindent
Solving the second equation for $g_{3}$ yields

\[ g_{3} = \frac{F_{4}(t)g_{2}(x,t)}{F_{3}(t)} \]

Considering the $O(1)$ equation next, we have the following system of equations

\begin{eqnarray}
&& g_{1x} = g_{4x} = 0 \\
&& f_{1t} - g_{2x} + f_{1}(g_{4} - g_{1}) = 0 \label{KDVFINAL2} \\
&& F_{3}(F_{4}f_{1})_{t} - f_{1}F_{4}F_{3t} - g_{2x}F_{4}F_{3} + F_{3}F_{4}f_{1}(g_{1} - g_{4}) = 0 \label{KDVFINAL3}
\end{eqnarray}

It follows that we must have $g_{1}(x,t) = G_{1}(t)$ and $g_{4}(x,t) = G_{4}(t)$ where $G_{1}$ and $G_{4}$ are arbitrary functions of $t$. Since $\eqref{KDVFINAL2}$ and $\eqref{KDVFINAL3}$ do not depend on the $a_{i}$ we will postpone solving them until the end. At this point the remaining conditions have been reduced to conditions involving soley the $a_{i}$ and the previously introduced arbitrary functions of $t$. The remaining conditions are given by

\begin{eqnarray}
&& \left(\frac{a_{5}}{a_{2}-3a_{1}}\right)_{x} + \left(\frac{a_{6}}{a_{2}-3a_{1}}\right)_{xxx} = 0 \label{KDVFINAL4} \\
&& \left(\frac{a_{3}}{a_{2}-3a_{1}}\right)_{x} = 0 \label{KDVFINAL5} \\
&& \left(\frac{a_{1}}{a_{2}-3a_{1}}\right)_{xx} = 0 \label{KDVFINAL6} \\
&& \left(\frac{a_{4}}{a_{2}-3a_{1}}\right)_{x} = 0 \label{KDVFINAL7}
\end{eqnarray}

One can easily solve the system of equations given by $\eqref{KDVFINAL1}$, $\eqref{KDVFINAL4}-\eqref{KDVFINAL7}$ yielding

\begin{eqnarray}
F_{4} &=& c_{1}F_{3}e^{2\int{(G_{4} - G_{1})dt}} \\
g_{2} &=& \int{(f_{1t}+f_{1}(G_{4}-G_{1}))dx} + F_{10} \\
a_{2} &=& -\frac{(3F_{1} - 1 - 3F_{2}x)a_{1}}{F_{2}x - F_{1}} \\
a_{3} &=& \frac{F_{5}a_{1}}{F_{2}x - F_{1}} \\
a_{4} &=& \frac{F_{6}a_{1}}{F_{2}x - F_{1}} \\
a_{6} &=& \frac{(F_{7} + F_{8}x + F_{9}x^{2})a_{1}}{F_{2}x - F_{1}} - \int^{x}{\int^{y}{\frac{a_{5}(z,t) \ dz\ dy}{a_{2}(z,t)-3a_{1}(z,t)}}} \\
a_{7} &=& \frac{a_{2}-3a_{1}}{F_{3}}\left(\frac{F_{3}}{a_{2}-3a_{1}}\right)_{t} + (a_{2}-3a_{1})\left(\frac{a_{8}}{a_{2}-3a_{1}}\right)_{x} + G_{4}-G_{1}
\end{eqnarray}

\noindent
where $F_{5-10}$ are arbitrary functions of $t$. Note that $a_{1},a_{5}$ and $a_{8}$ have no restrictions beyond the appropriate differentiability and integrability conditions. 

{\it The Lax pair for the generalized variable-coefficient KdV equation with the previous integrability conditions is therefore given by}

\begin{eqnarray}
F &=& \mathbb{X}_{1} + \mathbb{X}_{2}u \\
G &=& -a_{6}\mathbb{X}_{2}u_{xxxx} + (a_{6}\mathbb{X}_{2})_{x}u_{xxx} - \mathbb{X}_{2}(a_{1}u + a_{5})u_{xx} - (a_{6}\mathbb{X}_{2})_{xx}u_{xx} - a_{8}\mathbb{X}_{2}u \nonumber \\
&& + \frac{1}{2}a_{1}\mathbb{X}_{2}u_{x}^{2} - \frac{1}{2}a_{2}\mathbb{X}_{2}u_{x}^{2} + (a_{1}\mathbb{X}_{2})_{x}uu_{x} - \frac{1}{3}a_{3}\mathbb{X}_{2}u^{3} - \frac{1}{2}a_{4}\mathbb{X}_{2}u^{2} + \mathbb{X}_{0}
\end{eqnarray}

Next, we consider the modified KdV (MKdV) equation briefly as our third example.

\section{The Modified Korteweg-deVries (MKdV) Equation}

For this example we consider the mKdV equation given by

\begin{equation}
v_{t} + b_{1}v_{xxx} + b_{2}v^{2}v_{x} = 0 \label{MKDV}
\end{equation}

\noindent
where $b_{1}$ and $b_{2}$ are arbitrary functions of $x$ and $t$. Following the procedure we let 

\[ \mathbb{F} = \mathbb{F}(x,t,u) ,\ \ \ \mathbb{G} = \mathbb{G}(x,t,u,u_{x},u_{xx}) \]

Plugging this into $\eqref{ZCC}$ we obtain

\begin{equation}
\mathbb{F}_{t} + \mathbb{F}_{v}v_{t} - \mathbb{G}_{x} - \mathbb{G}_{v}v_{x} - \mathbb{G}_{v_{x}}v_{xx} - \mathbb{G}_{v_{xx}}v_{xxx} + [\mathbb{F},\mathbb{G}] = 0
\end{equation}

\noindent
Using $\eqref{MKDV}$ to substitute for $v_{t}$ we have

\begin{equation}
\mathbb{F}_{t} - \mathbb{G}_{x} - (\mathbb{G}_{v}+b_{2}\mathbb{F}_{v}v^{2})v_{x} - \mathbb{G}_{v_{x}}v_{xx} - (\mathbb{G}_{v_{xx}}+b_{1}\mathbb{F}_{v})v_{xxx} + [\mathbb{F},\mathbb{G}] = 0 \label{MKDVstep1}
\end{equation} 

\noindent
Since $\mathbb{F}$ and $\mathbb{G}$ do not depend on $v_{xxx}$ we can set the coefficient of the $v_{xxx}$ term to zero from which we have

\[ \mathbb{G}_{v_{xx}}+b_{1}\mathbb{F}_{v} = 0 \Rightarrow \mathbb{G} = -b_{1}\mathbb{F}_{v}v_{xx} + \mathbb{K}^{0}(x,t,v,v_{x}) \]

\noindent
Substituting this into $\eqref{MKDVstep1}$ we have

\begin{equation}
\mathbb{F}_{t} + (b_{1}\mathbb{F}_{v})_{x}v_{xx} - \mathbb{K}^{0}_{x} + b_{1}\mathbb{F}_{vv}v_{x}v_{xx} - \mathbb{K}^{0}_{v}v_{x} - b_{2}\mathbb{F}_{v}v^{2}v_{x} - \mathbb{K}^{0}_{v_{x}}v_{xx} - b_{1}[\mathbb{F},\mathbb{F}_{v}]v_{xx} + [\mathbb{F},\mathbb{K}^{0}] = 0 \label{MKDVstep2}
\end{equation}

Since $\mathbb{F}$ and $\mathbb{K}^{0}$ do not depend on $v_{xx}$ we can equate the coefficient of the $v_{xx}$ term to zero from which we require

\begin{equation}
(b_{1}\mathbb{F}_{v})_{x} + b_{1}\mathbb{F}_{vv}v_{x} - \mathbb{K}^{0}_{v_{x}} - b_{1}[\mathbb{F},\mathbb{F}_{v}] = 0
\end{equation}

\noindent
Solving for $\mathbb{K}^{0}$ we have

\[ \mathbb{K}^{0} = (b_{1}\mathbb{F}_{v})_{x}v_{x} + \frac{1}{2}b_{1}\mathbb{F}_{vv}v_{x}^{2} - b_{1}[\mathbb{F},\mathbb{F}_{v}]v_{x} + \mathbb{K}^{1}(x,t,v) \]

\noindent
Substituting this expression into $\eqref{MKDVstep2}$ we have

\begin{eqnarray}
&& \mathbb{F}_{t} - (b_{1}\mathbb{F}_{v})_{xx}v_{x} - \frac{1}{2}(b_{1}\mathbb{F}_{vv})_{x}v_{x}^{2} + (b_{1}[\mathbb{F},\mathbb{F}_{v}])_{x}v_{x} - \mathbb{K}^{1}_{x} - (b_{1}\mathbb{F}_{vv})_{x}v_{x}^{2} - \frac{1}{2}b_{1}\mathbb{F}_{vvv}v_{x}^{3} \nonumber \\
&& - \mathbb{K}^{1}_{v}v_{x} + b_{1}[\mathbb{F},\mathbb{F}_{vv}]v_{x}^{2} - b_{2}\mathbb{F}_{v}v^{2}v_{x} + [\mathbb{F},(b_{1}\mathbb{F}_{v})_{x}]v_{x} + \frac{1}{2}b_{1}[\mathbb{F},\mathbb{F}_{vv}]v_{x}^{2} - b_{1}[\mathbb{F},[\mathbb{F},\mathbb{F}_{v}]]v_{x} \nonumber \\
&& + [\mathbb{F},\mathbb{K}^{1}] = 0 \label{MKDVstep3}
\end{eqnarray}

Since $\mathbb{F}$ and $\mathbb{K}^{1}$ do not depend on $v_{x}$ we can equate the coefficients of the $v_{x}$, $v_{x}^{2}$ and $v_{x}^{3}$ terms to zero from which we obtain the system

\begin{eqnarray}
O(v_{x}^{3}) &:& \mathbb{F}_{vvv} = 0 \\
O(v_{x}^{2}) &:&  \frac{1}{2}(b_{1}\mathbb{F}_{vv})_{x} + (b_{1}\mathbb{F}_{vv})_{x} - b_{1}[\mathbb{F},\mathbb{F}_{vv}] - \frac{1}{2}b_{1}[\mathbb{F},\mathbb{F}_{vv}] = 0 \\
O(v_{x}) &:&  (b_{1}\mathbb{F}_{v})_{xx} - (b_{1}[\mathbb{F},\mathbb{F}_{v}])_{x} + \mathbb{K}^{1}_{v} + b_{2}\mathbb{F}_{v}v^{2} - [\mathbb{F},(b_{1}\mathbb{F}_{v})_{x}] + b_{1}[\mathbb{F},[\mathbb{F},\mathbb{F}_{v}]] = 0
\end{eqnarray}

Since the MKdV equation does not contain a $vv_{t}$ term and for ease of computation we take require $\mathbb{F}_{vv} = 0$ from which we have $\mathbb{F} = \mathbb{X}_{1}(x,t) + \mathbb{X}_{2}(x,t)v$. For the $O(v_{x})$ equation we solve for $\mathbb{K}^{1}$ and thus have

\begin{eqnarray}
\mathbb{K}^{1} &=& -(b_{1}\mathbb{X}_{2})_{xx}v + (b_{1}[\mathbb{X}_{1},\mathbb{X}_{2}])_{x}v + [\mathbb{X}_{1},(b_{1}\mathbb{X}_{2})_{x}]v + \frac{1}{2}[\mathbb{X}_{2},(b_{1}\mathbb{X}_{2})_{x}]v^{2} - b_{1}[\mathbb{X}_{1},[\mathbb{X}_{1},\mathbb{X}_{2}]]v \nonumber \\
&& - \frac{1}{3}b_{2}\mathbb{X}_{2}v^{3} - \frac{1}{2}b_{1}[\mathbb{X}_{2},[\mathbb{X}_{1},\mathbb{X}_{2}]]v^{2} + \mathbb{X}_{0}(x,t)
\end{eqnarray}

\noindent
Substituting this expression for $\mathbb{K}^{1}$ into $\eqref{MKDVstep3}$ we obtain

\begin{eqnarray}
&& \mathbb{X}_{1,t} + (b_{1}\mathbb{X}_{2})_{xxx}v - (b_{1}[\mathbb{X}_{1},\mathbb{X}_{2}])_{xx}v + \frac{1}{3}(b_{2}\mathbb{X}_{2})_{x}v^{3} - ([\mathbb{X}_{1},(b_{1}\mathbb{X}_{2})_{x}])_{x}v - \frac{1}{2}([\mathbb{X}_{2},(b_{1}\mathbb{X}_{2})_{x}])_{x}v^{2} \nonumber \\
&& + (b_{1}[\mathbb{X}_{1},[\mathbb{X}_{1},\mathbb{X}_{2}]])_{x}v + \frac{1}{2}(b_{1}[\mathbb{X}_{2},[\mathbb{X}_{1},\mathbb{X}_{2}]])_{x}v^{2} - \mathbb{X}_{0,x} - [\mathbb{X}_{1},(b_{1}\mathbb{X}_{2})_{xx}]v + [\mathbb{X}_{1},(b_{1}[\mathbb{X}_{1},\mathbb{X}_{2}])_{x}]v \nonumber \\
&& + \mathbb{X}_{2,t}v - \frac{1}{3}b_{2}[\mathbb{X}_{1},\mathbb{X}_{2}]v^{3} + [\mathbb{X}_{1},[\mathbb{X}_{1},(b_{1}\mathbb{X}_{2})_{x}]]v + \frac{1}{2}[\mathbb{X}_{1},[\mathbb{X}_{2},(b_{1}\mathbb{X}_{2})_{x}]]v^{2} - b_{1}[\mathbb{X}_{1},[\mathbb{X}_{1},[\mathbb{X}_{1},\mathbb{X}_{2}]]]v \nonumber \\
&& - \frac{1}{2}b_{1}[\mathbb{X}_{1},[\mathbb{X}_{2},[\mathbb{X}_{1},\mathbb{X}_{2}]]]v^{2} - [\mathbb{X}_{2},(b_{1}\mathbb{X}_{2})_{xx}]v^{2} + [\mathbb{X}_{2},(b_{1}[\mathbb{X}_{1},\mathbb{X}_{2}])_{x}]v^{2} + [\mathbb{X}_{2},[\mathbb{X}_{1},(b_{1}\mathbb{X}_{2})_{x}]]v^{2} \nonumber \\
&& + [\mathbb{X}_{1},\mathbb{X}_{0}] + \frac{1}{2}[\mathbb{X}_{2},[\mathbb{X}_{2},(b_{1}\mathbb{X}_{2})_{x}]]v^{3} - b_{1}[\mathbb{X}_{2},[\mathbb{X}_{1},[\mathbb{X}_{1},\mathbb{X}_{2}]]]v^{2} - \frac{1}{2}b_{1}[\mathbb{X}_{2},[\mathbb{X}_{2},[\mathbb{X}_{1},\mathbb{X}_{2}]]]v^{3} \nonumber \\
&& + [\mathbb{X}_{2},\mathbb{X}_{0}]v = 0
\end{eqnarray}

Since the $\mathbb{X}_{i}$ do not depend on $v$ we can equate the coefficients of the different powers of $v$ to zero. We thus obtain the constraints

\begin{eqnarray}
O(1) &:& \mathbb{X}_{1,t} - \mathbb{X}_{0,x} + [\mathbb{X}_{1},\mathbb{X}_{0}] \\
O(v) &:& \mathbb{X}_{2,t} - ([\mathbb{X}_{1},(b_{1}\mathbb{X}_{2})_{x}])_{x} + (b_{1}[\mathbb{X}_{1},[\mathbb{X}_{1},\mathbb{X}_{2}]])_{x} - [\mathbb{X}_{1},(b_{1}\mathbb{X}_{2})_{xx}] + [\mathbb{X}_{1},(b_{1}[\mathbb{X}_{1},\mathbb{X}_{2}])_{x}] \nonumber \\
&& - (b_{1}[\mathbb{X}_{1},\mathbb{X}_{2}])_{xx} + (b_{1}\mathbb{X}_{2})_{xxx} + [\mathbb{X}_{1},[\mathbb{X}_{1},(b_{1}\mathbb{X}_{2})_{x}]] - b_{1}[\mathbb{X}_{1},[\mathbb{X}_{1},[\mathbb{X}_{1},\mathbb{X}_{2}]]] \nonumber \\
&& + [\mathbb{X}_{2},\mathbb{X}_{0}] = 0 \\
O(v^{2}) &:&  - \frac{1}{2}([\mathbb{X}_{2},(b_{1}\mathbb{X}_{2})_{x}])_{x} + \frac{1}{2}(b_{1}[\mathbb{X}_{2},[\mathbb{X}_{1},\mathbb{X}_{2}]])_{x} + \frac{1}{2}[\mathbb{X}_{1},[\mathbb{X}_{2},(b_{1}\mathbb{X}_{2})_{x}]] - [\mathbb{X}_{2},(b_{1}\mathbb{X}_{2})_{xx}] \nonumber \\
&& - \frac{1}{2}b_{1}[\mathbb{X}_{1},[\mathbb{X}_{2},[\mathbb{X}_{1},\mathbb{X}_{2}]]] + [\mathbb{X}_{2},(b_{1}[\mathbb{X}_{1},\mathbb{X}_{2}])_{x}] - b_{1}[\mathbb{X}_{2},[\mathbb{X}_{1},[\mathbb{X}_{1},\mathbb{X}_{2}]]] \nonumber \\
&& + [\mathbb{X}_{2},[\mathbb{X}_{1},(b_{1}\mathbb{X}_{2})_{x}]] = 0 \\
O(v^{3}) &:& \frac{1}{3}(b_{2}\mathbb{X}_{2})_{x} - \frac{1}{3}b_{2}[\mathbb{X}_{1},\mathbb{X}_{2}] + \frac{1}{2}[\mathbb{X}_{2},[\mathbb{X}_{2},(b_{1}\mathbb{X}_{2})_{x}]] - \frac{1}{2}b_{1}[\mathbb{X}_{2},[\mathbb{X}_{2},[\mathbb{X}_{1},\mathbb{X}_{2}]]] = 0
\end{eqnarray}

\noindent
Note that if we decouple the $O(v^{3})$ equation into the following equations

\begin{eqnarray}
&& b_{1}[\mathbb{X}_{1},\mathbb{X}_{2}] - (b_{1}\mathbb{X}_{2})_{x} = 0 \\
&& b_{2}[\mathbb{X}_{1},\mathbb{X}_{2}] - (b_{2}\mathbb{X}_{2})_{x} = 0
\end{eqnarray}

\noindent
we find that the $O(v^{2})$ equation is immediately satisfied and the $O(v)$ equation reduces to

\begin{equation}
[\mathbb{X}_{2},\mathbb{X}_{0}] + \mathbb{X}_{2,t} = 0
\end{equation}

{\it Again we should note that had we opted instead for the forms}

\[ \mathbb{X}_{0} = \begin{bmatrix}
g_{1}(x,t) & g_{4}(x,t) \\
g_{10}(x,t) & g_{16}(x,t)
\end{bmatrix}, \ \ \mathbb{X}_{1} = \begin{bmatrix}
f_{1}(x,t) & f_{3}(x,t) \\
f_{5}(x,t) & f_{7}(x,t)
\end{bmatrix}, \ \ \mathbb{X}_{2} = \begin{bmatrix}
f_{2}(x,t) & f_{4}(x,t) \\
f_{6}(x,t) & f_{8}(x,t)
\end{bmatrix} \]

\noindent
{\it we would obtain an equivalent system to that obtained in $\cite{Lecce}$ for the mKdV. The additional unknown functions which appear in Khawaja's method ($\cite{Lecce}$) can again be introduced with the proper substitutions via their functional dependence on the twelve unknown functions given above.}

Therefore utilizing the same generators as in the generalized KdV equation we obtain the system of equations

\begin{eqnarray}
&& f_{1}f_{4} - f_{2}f_{3} = 0 \\
&& (b_{1}f_{j})_{x} = (b_{2}f_{j})_{x} = 0, \ \ \ j=3,4 \\
&& f_{3}g_{3} - f_{4}g_{2} = 0 \\
&& f_{jt} + (-1)^{j}f_{j}(g_{1} - g_{4}) = 0, \ \ \ j=3,4 \\
&& g_{jx} + (-1)^{j}(f_{1}g_{3} - f_{2}g_{2}) = 0, \ \ \ j=1,4 \\
&& f_{jt} - g_{(j+1)x} + (-1)^{j}f_{j}(g_{4} - g_{1}) = 0, \ \ \ j=2,3
\end{eqnarray}

Solving this system yields the following

\begin{eqnarray}
f_{j}(x,t) &=& \frac{F_{j}(t)}{b_{1}(x,t)}, \ \ \ j=3,4 \\
g_{3}(x,t) &=& \frac{F_{4}(t)g_{2}(x,t)}{F_{3}(t)} \\
f_{2}(x,t) &=& \frac{F_{4}(t)f_{1}(x,t)}{F_{3}(t)} \\
g_{1}(x,t) &=& G_{1}(t) \\
g_{4}(x,t) &=& G_{4}(t)
\end{eqnarray}

Subject to the constraints

\begin{eqnarray}
&& \left(\frac{F_{j}}{b_{1}}\right)_{t} + (-1)^{j}(G_{1} - G_{4}) = 0, \ \ \ j=3,4 \label{MKDVFINAL1} \\
&& \left(\frac{b_{2}F_{j}}{b_{1}}\right)_{x} = 0 \ \ \ j=3,4 \label{MKDVFINAL2} \\
&& f_{jt} - g_{(j+1)x} + (-1)^{j}f_{j}(G_{4} - G_{1}) = 0, \ \ \ j=2,3
\end{eqnarray}

Solving $\eqref{MKDVFINAL1}$ and $\eqref{MKDVFINAL2}$ for $F_{4},b_{1},b_{2}$ and $G_{4}$ we obtain

\begin{eqnarray}
F_{4}(t) &=& \frac{c_{1}F_{2}(t)^{2}}{F_{3}(t)} \\
b_{1}(x,t) &=& F_{1}(x)F_{2}(t) \\
b_{2}(x,t) &=& F_{1}(x)F_{5}(t) \\
G_{4}(t) &=& \frac{F_{3}(t)F_{2}'(t) - F_{2}(t)F_{3}'(t) + G_{1}(t)F_{2}(t)F_{3}(t)}{F_{2}(t)F_{3}(t)} \\
g_{2}(x,t) &=& \int{\frac{[f_{1}(x,t)]_{t}F_{3}(t)F_{2}(t) - f_{1}(x,t)F_{3}'(t)F_{2}(t) + f_{1}(x,t)F_{3}(t)F_{2}'(t)}{F_{3}(t)F_{2}(t)}dx} + F_{6}(t)
\end{eqnarray}

\noindent
where $F_{1}$ and $F_{2}$ are arbitrary functions in their respective variables and $c_{1}$ is an arbitrary constant. 

{\it The Lax pair for the variable-coefficient MKdV equation with the previous integrability conditions is thus given by}

\begin{eqnarray}
F &=& \mathbb{X}_{1} + \mathbb{X}_{2}v \\
G &=& -b_{1}\mathbb{X}_{2}v_{xx} - \frac{1}{3}b_{2}\mathbb{X}_{2}v^{3} + \mathbb{X}_{0}
\end{eqnarray}

Next, as our final example, we consider the Derivative Nonlinear Schrodinger (DNLS) Equation.

\section{The Derivative Nonlinear Schrodinger (DNLS) Equation}

Consider the derivative NLS (DNLS) given by the system

\begin{subequations} 
\begin{align}
        iq_{t} + a_{1}q_{xx} + 2iqrq_{x} + ia_{2}q^{2}r_{x}&=0,  \label{DNLSa} \\
        -ir_{t} + a_{1}r_{xx} - 2ia_{2}rq_{x}r - ia_{2}r^{2}q_{x}&=0, \label{DNLSb}
\end{align}
\end{subequations}

\noindent
where $a_{1}$ and $a_{2}$ are arbitrary functions of $x$ and $t$. The details of this example will be similar to that of the standard NLS. Following the procedure we let

\[ \mathbb{F} = \mathbb{F}(x,t,r,q) ,\ \ \ \mathbb{G} = \mathbb{G}(x,t,r,q,r_{x},q_{x}) \]

Plugging this into $\eqref{ZCC}$ we obtain

\begin{equation}
\mathbb{F}_{t} + \mathbb{F}_{q}q_{t} + \mathbb{F}_{r}r_{t} - \mathbb{G}_{x} - \mathbb{G}_{q}q_{x} - \mathbb{G}_{q_{x}}q_{xx} - \mathbb{G}_{r}r_{x} - \mathbb{G}_{r_{x}}r_{xx} + [\mathbb{F},\mathbb{G}] = 0
\end{equation}

\noindent
Now substituting for $q_{t}$ and $r_{t}$ using $\eqref{DNLSa}$ and $\eqref{DNLSb}$ we have

\begin{eqnarray}
&& \mathbb{F}_{t} + (ia_{1}\mathbb{F}_{q}-\mathbb{G}_{q_{x}})q_{xx} - (ia_{1}\mathbb{F}_{r}+\mathbb{G}_{r_{x}})r_{xx} - \mathbb{F}_{q}(2a_{2}rqq_{x} + a_{2}q^{2}r_{x}) \nonumber \\
&& - \mathbb{F}_{r}(2a_{2}qrr_{x} + a_{2}r^{2}q_{x}) - \mathbb{G}_{x} - \mathbb{G}_{q}q_{x} - \mathbb{G}_{r}r_{x} + [\mathbb{F},\mathbb{G}] = 0 \label{DNLSP1}
\end{eqnarray}

Since $\mathbb{F}$ and $\mathbb{G}$ do not depend on $q_{xx}$ or $r_{xx}$ we can set the coefficients of the $q_{xx}$ and $r_{xx}$ terms to zero. This requires

\begin{equation}
ia_{1}\mathbb{F}_{q} - \mathbb{G}_{q_{x}} = 0, \ \mbox{and} \ ia_{1}\mathbb{F}_{r} + \mathbb{G}_{r_{x}} = 0
\end{equation}

\noindent
Solving this in the same manner as in the NLS example we obtain

\begin{equation}
G = ia_{1}(\mathbb{F}_{q}q_{x} - \mathbb{F}_{r}r_{x}) + \mathbb{K}^{0}(x,t,r,q)
\end{equation}

\noindent
Plugging this into $\eqref{DNLSP1}$ we obtain

\begin{eqnarray}
&& \mathbb{F}_{t} - \mathbb{F}_{q}(2a_{2}rqq_{x} + a_{2}q^{2}r_{x}) - \mathbb{F}_{r}(2a_{2}qrr_{x} + a_{2}r^{2}q_{x}) - i(a_{1}\mathbb{F}_{q})_{x}q_{x} + i(a_{1}\mathbb{F}_{r})_{x}r_{x} - \mathbb{K}^{0}_{x} \nonumber \\
&& + ia_{1}\mathbb{F}_{rr}r_{x}^{2} - \mathbb{K}^{0}_{r}r_{x} - ia_{1}\mathbb{F}_{qq}q_{x}^{2} - \mathbb{K}^{0}_{q}q_{x} + ia_{1}[\mathbb{F},\mathbb{F}_{q}]q_{x} - ia_{1}[\mathbb{F},\mathbb{F}_{r}]r_{x} + [\mathbb{F},\mathbb{K}^{0}] = 0 \label{DNLSP2}
\end{eqnarray}

Now since $\mathbb{F}$ and $\mathbb{K}^{0}$ do not depend on $q_{x}$ or $r_{x}$ we can set the coefficients of the different powers of $r_{x}$ and $q_{x}$ to zero. Thus, setting the coefficients of the $q_{x}^{2}$ and $r_{x}^{2}$ terms to zero we have

\begin{equation}
-ia_{1}\mathbb{F}_{qq} = ia_{1}\mathbb{F}_{rr} = 0
\end{equation}

\noindent
from which it follows $\mathbb{F} = \mathbb{X}_{1}(x,t) + \mathbb{X}_{2}(x,t)r + \mathbb{X}_{3}(x,t)q + \mathbb{X}_{4}(x,t)qr$. Now setting the coefficients of the $q_{x}$ and $r_{x}$ terms to zero we have

\begin{eqnarray}
&& -a_{2}\mathbb{X}_{3}q^{2} - a_{2}\mathbb{X}_{4}q^{2}r + i(a_{1}\mathbb{X}_{2})_{x} + i(a_{1}\mathbb{X}_{4})_{x}q - \mathbb{K}^{0}_{r} - ia_{1}[\mathbb{X}_{1},\mathbb{X}_{2}] - ia_{1}[\mathbb{X}_{1},\mathbb{X}_{4}]q \nonumber \\
&& - ia_{1}[\mathbb{X}_{3},\mathbb{X}_{2}]q - ia_{1}[\mathbb{X}_{3},\mathbb{X}_{4}]q^{2} - 2a_{2}\mathbb{X}_{2}rq - 2a_{2}\mathbb{X}_{4}q^{2}r = 0 \label{XIEQ2} \\
&& -a_{2}\mathbb{X}_{2}r^{2} - a_{2}\mathbb{X}_{4}r^{2}q - i(a_{1}\mathbb{X}_{3})_{x} - i(a_{1}\mathbb{X}_{4})_{x}r - \mathbb{K}^{0}_{q} + ia_{1}[\mathbb{X}_{1},\mathbb{X}_{3}] + ia_{1}[\mathbb{X}_{1},\mathbb{X}_{4}]r \nonumber \\
&& + ia_{1}[\mathbb{X}_{2},\mathbb{X}_{3}]r + ia_{1}[\mathbb{X}_{2},\mathbb{X}_{4}]r^{2} - 2a_{2}\mathbb{X}_{3}rq - 2a_{2}\mathbb{X}_{4}r^{2}q = 0 \label{ETA2}
\end{eqnarray}

In much the same way as for the NLS we denote the left-hand side of $\eqref{XIEQ2}$ as $\xi(r,q)$ and the left-hand side of $\eqref{ETA2}$ as $\eta(r,q)$. For recovery of $\mathbb{K}^{0}$ we require that $\xi_{q} = \eta_{r}$. Thus, computing $\xi_{q}$ and $\eta_{r}$ we find

\begin{eqnarray}
\xi_{q} &=& -2a_{2}\mathbb{X}_{3}q - 2a_{2}\mathbb{X}_{4}qr + i(a_{1}\mathbb{X}_{4})_{x} - ia_{1}[\mathbb{X}_{1},\mathbb{X}_{4}] - ia_{1}[\mathbb{X}_{3},\mathbb{X}_{2}] - 2ia_{1}[\mathbb{X}_{3},\mathbb{X}_{4}]q \nonumber \\
&& - 2a_{2}\mathbb{X}_{2}r - 4a_{2}\mathbb{X}_{4}qr \\
\eta_{r} &=&  -2a_{2}\mathbb{X}_{2}r - 2a_{2}\mathbb{X}_{4}rq - i(a_{1}\mathbb{X}_{4})_{x} + ia_{1}[\mathbb{X}_{1},\mathbb{X}_{4}] + ia_{1}[\mathbb{X}_{2},\mathbb{X}_{3}] + 2ia_{1}[\mathbb{X}_{2},\mathbb{X}_{4}]r \nonumber \\
&& - 2a_{2}\mathbb{X}_{3}q - 4a_{2}\mathbb{X}_{4}rq
\end{eqnarray}

\noindent
from which it follows that we must have

\begin{equation}
2i(a_{1}\mathbb{X}_{4})_{x} - 2ia_{1}[\mathbb{X}_{1},\mathbb{X}_{4}] - 2ia_{1}[\mathbb{X}_{3},\mathbb{X}_{4}]q - 2ia_{1}[\mathbb{X}_{2},\mathbb{X}_{4}]r = 0
\end{equation}

Since the $\mathbb{X}_{i}$ do not depend on $q$ or $r$ this previous condition requires

\begin{eqnarray}
&& 2i(a_{1}\mathbb{X}_{4})_{x} - 2ia_{1}[\mathbb{X}_{1},\mathbb{X}_{4}] = 0 \\
&& - 2ia_{1}[\mathbb{X}_{3},\mathbb{X}_{4}] = 0 \\
&& - 2ia_{1}[\mathbb{X}_{2},\mathbb{X}_{4}] = 0
\end{eqnarray}

As with the standard NLS we will take $\mathbb{X}_{4} = 0$ in order to simplify computations. Therefore

\begin{eqnarray}
\mathbb{K}^{0}_{q} &=& -a_{2}\mathbb{X}_{2}r^{2} - i(a_{1}\mathbb{X}_{3})_{x} + ia_{1}[\mathbb{X}_{1},\mathbb{X}_{3}] + ia_{1}[\mathbb{X}_{2},\mathbb{X}_{3}]r - 2a_{2}\mathbb{X}_{3}rq \\
\mathbb{K}^{0}_{r} &=& -a_{2}\mathbb{X}_{3}q^{2} + i(a_{1}\mathbb{X}_{2})_{x} - ia_{1}[\mathbb{X}_{1},\mathbb{X}_{2}] - ia_{1}[\mathbb{X}_{3},\mathbb{X}_{2}]q - 2a_{2}\mathbb{X}_{2}rq
\end{eqnarray}

Integrating the first equation with respect to $q$ yields

\[ \mathbb{K}^{0} = -a_{2}\mathbb{X}_{2}r^{2}q - i(a_{1}\mathbb{X}_{3})_{x}q + ia_{1}[\mathbb{X}_{1},\mathbb{X}_{3}]q + ia_{1}[\mathbb{X}_{2},\mathbb{X}_{3}]rq - a_{2}\mathbb{X}_{3}q^{2}r + \mathbb{K}^{*}(x,t,r) \]

Now differentiating this equation with respect to $r$ and mandating that it equal our previous expression for $\mathbb{K}^{0}_{r}$ we find that $\mathbb{K}^{*}$ must satisfy

\[ \mathbb{K}^{*}_{r} = i(a_{1}\mathbb{X}_{2})_{x} - ia_{1}[\mathbb{X}_{1},\mathbb{X}_{2}] \]

from which it follows

\[ \mathbb{K}^{*} = i(a_{1}\mathbb{X}_{2})_{x}r - ia_{1}[\mathbb{X}_{1},\mathbb{X}_{2}]r + \mathbb{X}_{0}(x,t) \]

\noindent
and thus

\begin{eqnarray}
\mathbb{K}^{0} &=& i(a_{1}\mathbb{X}_{2})_{x}r - i(a_{1}\mathbb{X}_{3})_{x}q - ia_{1}[\mathbb{X}_{1},\mathbb{X}_{2}]r + ia_{1}[\mathbb{X}_{1},\mathbb{X}_{3}]q + ia_{1}[\mathbb{X}_{2},\mathbb{X}_{3}]rq - a_{2}\mathbb{X}_{2}r^{2}q \nonumber \\
&& - a_{2}\mathbb{X}_{3}q^{2}r + \mathbb{X}_{0}(x,t)
\end{eqnarray}

Now plugging this and our expression for $\mathbb{F}$ into $\eqref{DNLSP2}$ we get

\begin{eqnarray}
&& \mathbb{X}_{1,t} + \mathbb{X}_{2,t}r + \mathbb{X}_{3,t}q - i(a_{1}\mathbb{X}_{2})_{xx}r + i(a_{1}\mathbb{X}_{3})_{xx}q + i(a_{1}[\mathbb{X}_{1},\mathbb{X}_{2}])_{x}r - i(a_{1}[\mathbb{X}_{1},\mathbb{X}_{3}])_{x}q \nonumber \\
&& - i(a_{1}[\mathbb{X}_{2},\mathbb{X}_{3}])_{x}rq + (a_{2}\mathbb{X}_{2})_{x}r^{2}q + (a_{2}\mathbb{X}_{3})_{x}q^{2}r - \mathbb{X}_{0,x} + i[\mathbb{X}_{1},(a_{1}\mathbb{X}_{2})_{x}]r - [\mathbb{X}_{1},(a_{1}\mathbb{X}_{3})_{x}]q \nonumber \\
&& - ia_{1}[\mathbb{X}_{1},[\mathbb{X}_{1},\mathbb{X}_{2}]]r + ia_{1}[\mathbb{X}_{1},[\mathbb{X}_{1},\mathbb{X}_{3}]]q + ia_{1}[\mathbb{X}_{1},[\mathbb{X}_{2},\mathbb{X}_{3}]]rq - a_{2}[\mathbb{X}_{1},\mathbb{X}_{2}]r^{2}q - a_{2}[\mathbb{X}_{1},\mathbb{X}_{3}]q^{2}r \nonumber \\
&& + [\mathbb{X}_{1},\mathbb{X}_{0}] + i[\mathbb{X}_{2},(a_{1}\mathbb{X}_{2})_{x}]r^{2} - i[\mathbb{X}_{2},(a_{1}\mathbb{X}_{3})_{x}]rq - ia_{1}[\mathbb{X}_{2},[\mathbb{X}_{1},\mathbb{X}_{2}]]r^{2} + ia_{1}[\mathbb{X}_{2},[\mathbb{X}_{1},\mathbb{X}_{3}]]rq \nonumber \\
&& + ia_{1}[\mathbb{X}_{2},[\mathbb{X}_{2},\mathbb{X}_{3}]]r^{2}q + [\mathbb{X}_{2},\mathbb{X}_{0}]r + i[\mathbb{X}_{3},(a_{1}\mathbb{X}_{2})_{x}]rq - i[\mathbb{X}_{3},(a_{1}\mathbb{X}_{3})_{x}]q^{2} - ia_{1}[\mathbb{X}_{3},[\mathbb{X}_{1},\mathbb{X}_{2}]]rq \nonumber \\
&& + ia_{1}[\mathbb{X}_{3},[\mathbb{X}_{1},\mathbb{X}_{3}]]q^{2} + ia_{1}[\mathbb{X}_{3},[\mathbb{X}_{2},\mathbb{X}_{3}]]q^{2}r + [\mathbb{X}_{3},\mathbb{X}_{0}]q = 0
\end{eqnarray}

Since the $\mathbb{X}_{i}$ are independent of $r$ and $q$ we equate the coefficients of the different powers of $r$ and $q$ to zero and thus obtain the following constraints:

\begin{eqnarray}
O(1) &:& \mathbb{X}_{1,t} - \mathbb{X}_{0,x} + [\mathbb{X}_{1},\mathbb{X}_{0}] = 0 \label{DNLSCOND1} \\
O(q) &:& \mathbb{X}_{3,t} + i(a_{1}\mathbb{X}_{3})_{xx} - i(a_{1}[\mathbb{X}_{1},\mathbb{X}_{3}])_{x} - i[\mathbb{X}_{1},(a_{1}\mathbb{X}_{3})_{x}] + ia_{1}[\mathbb{X}_{1},[\mathbb{X}_{1},\mathbb{X}_{3}]] \nonumber \\
&& + [\mathbb{X}_{3},\mathbb{X}_{0}] = 0 \label{DNLSCOND2} \\
O(r) &:& \mathbb{X}_{2,t} - i(a_{1}\mathbb{X}_{2})_{xx} + i(a_{1}[\mathbb{X}_{1},\mathbb{X}_{2}])_{x} + i[\mathbb{X}_{1},(a_{1}\mathbb{X}_{2})_{x}] - ia_{1}[\mathbb{X}_{1},[\mathbb{X}_{1},\mathbb{X}_{2}]] \nonumber \\
&& + [\mathbb{X}_{2},\mathbb{X}_{0}] = 0 \label{DNLSCOND3} \\
O(rq) &:& - (a_{1}[\mathbb{X}_{2},\mathbb{X}_{3}])_{x} + a_{1}[\mathbb{X}_{1},[\mathbb{X}_{2},\mathbb{X}_{3}]] - [\mathbb{X}_{2},(a_{1}\mathbb{X}_{3})_{x}] + a_{1}[\mathbb{X}_{2},[\mathbb{X}_{1},\mathbb{X}_{3}]] + [\mathbb{X}_{3},(a_{1}\mathbb{X}_{2})_{x}] \nonumber \\
&&  - a_{1}[\mathbb{X}_{3},[\mathbb{X}_{1},\mathbb{X}_{2}]] = 0 \label{DNLSCOND4} \\
O(q^{2}) &:& - [\mathbb{X}_{3},(a_{1}\mathbb{X}_{3})_{x}] + a_{1}[\mathbb{X}_{3},[\mathbb{X}_{1},\mathbb{X}_{3}]] = 0 \label{DNLSCOND5} \\
O(r^{2}) &:& [\mathbb{X}_{2},(a_{1}\mathbb{X}_{2})_{x}] - a_{1}[\mathbb{X}_{2},[\mathbb{X}_{1},\mathbb{X}_{2}]] = 0 \label{DNLSCOND6} \\
O(r^{2}q) &:& (a_{2}\mathbb{X}_{2})_{x} - a_{2}[\mathbb{X}_{1},\mathbb{X}_{2}] + ia_{1}[\mathbb{X}_{2},[\mathbb{X}_{2},\mathbb{X}_{3}]] = 0 \label{DNLSCOND7} \\
O(q^{2}r) &:& (a_{2}\mathbb{X}_{3})_{x} - a_{2}[\mathbb{X}_{1},\mathbb{X}_{3}] + ia_{1}[\mathbb{X}_{3},[\mathbb{X}_{2},\mathbb{X}_{3}]] = 0 \label{DNLSCOND8}
\end{eqnarray}

Allowing the following forms for the generators

\begin{equation}
\mathbb{X}_{0} = \begin{bmatrix}
g_{1} & g_{2} \\
g_{3} & g_{4}
\end{bmatrix}, \ \ \mathbb{X}_{1} = \begin{bmatrix}
f_{1} & 0 \\
0 & f_{2}
\end{bmatrix}, \ \ \mathbb{X}_{2} = \begin{bmatrix}
0 & f_{3} \\
0 & 0
\end{bmatrix}, \ \ \mathbb{X}_{3} = \begin{bmatrix}
0 & 0 \\
f_{4} & 0
\end{bmatrix}
\end{equation}

Note that with this choice the $\eqref{DNLSCOND5}$ and $\eqref{DNLSCOND6}$ equations are immediately satisfed. From $\eqref{DNLSCOND2}$ and $\eqref{DNLSCOND3}$ we obtain the conditions

\begin{eqnarray}
&& g_{3}f_{4} = g_{2}f_{3} = 0 \\
&& f_{4t} + i(a_{1}f_{4})_{xx} - i(a_{1}f_{4}(f_{1} - f_{2}))_{x} + (f_{2} - f_{1})(a_{1}f_{4})_{x} + ia_{1}f_{4}(f_{1} - f_{2})^{2} \nonumber \\
&& + f_{4}(g_{4} - g_{1}) = 0 \label{DNLSFINAL2} \\
&& f_{3t} - i(a_{1}f_{3})_{xx} - i(a_{1}f_{3}(f_{1} - f_{2}))_{x} + (f_{2} - f_{1})(a_{1}f_{3})_{x} - ia_{1}f_{3}(f_{1} - f_{2})^{2} \nonumber \\
&& - f_{3}(g_{4} - g_{1}) = 0 \label{DNLSFINAL3}
\end{eqnarray}

To keep $\mathbb{X}_{2}$ and $\mathbb{X}_{3}$ nonzero we force $g_{2} = g_{3} = 0$. The condition given by $\eqref{DNLSCOND4}$ becomes the single equation

\begin{equation}
(a_{1}f_{3}f_{4})_{x} + f_{3}(a_{1}f_{4})_{x} + f_{4}(a_{1}f_{3})_{x} = 0 \label{DNLSFINAL4} \\
\end{equation}

The final two conditions now yield the system

\begin{eqnarray}
&& (a_{2}f_{3})_{x} - a_{2}f_{3}(f_{2} - f_{1}) - 2ia_{1}f_{3}^{2}f_{4} = 0 \label{DNLSFINAL5} \\
&& (a_{2}f_{4})_{x} + a_{2}f_{4}(f_{2} - f_{1}) + 2ia_{1}f_{4}^{2}f_{3} = 0 \label{DNLSFINAL6}
\end{eqnarray}

At this point solution of the system given by $\eqref{DNLSCOND1}$ and $\eqref{DNLSFINAL2} - \eqref{DNLSFINAL6}$ such that the $a_{i}$ are real-valued requires either $f_{3} = 0$ or $f_{4} = 0$. Without loss of generality we choose $f_{3} = 0$ from which we obtain the new system of equations

\begin{eqnarray}
&& f_{1t} - g_{1x} = 0 \label{DNLSE1} \\
&& f_{2t} - g_{4x} = 0 \label{DNLSE2} \\
&& f_{4t} + i(a_{1}f_{4})_{xx} - i(a_{1}f_{4}(f_{1} - f_{2}))_{x} + (f_{2} - f_{1})(a_{1}f_{4})_{x} + ia_{1}f_{4}(f_{1} - f_{2})^{2} \nonumber \\
&& + f_{4}(g_{4} - g_{1}) = 0 \label{DNLSE3} \\
&& (a_{2}f_{4})_{x} + a_{2}f_{4}(f_{2} - f_{1}) = 0 \label{DNLSE4}
\end{eqnarray}

Solving $\eqref{DNLSE1},\eqref{DNLSE3}$ and $\eqref{DNLSE4}$ for $f_{1},g_{4}$ and $f_{2}$, respectively we obtain

\begin{eqnarray}
&& f_{1} = \int{g_{1x}dt} + F_{1}(x) \\
&& f_{2} = -\frac{(a_{2}f_{4})_{x}}{a_{2}f_{4}} + \int{g_{1x}dt} + F_{1}(x) \\
&& g_{4} = \frac{-ia_{2}^{2}f_{4}a_{1xx} + ia_{1}a_{2}f_{4}a_{2xx} - 2ia_{1}a_{2x}^{2}f_{4} + 2ia_{2}a_{2x}a_{1x}f_{4} - f_{4t}a_{2}^{2}}{a_{2}^{2}f_{4}} + g_{1}
\end{eqnarray}

Plugging this into $\eqref{DNLSE2}$ yields the integrability condition

\begin{eqnarray}
&& a_{2}^{3}a_{1xxx} - ia_{2t}a_{2x}a_{2} + ia_{2xt}a_{2}^{2} - 3a_{2}^{2}a_{2xx}a_{1x} - 4a_{2x}^{3}a_{1} + 5a_{1}a_{2}a_{2x}a_{2xx} + 4a_{2x}^{2}a_{2}a_{1x} \nonumber \\
&& - a_{2}^{2}a_{1}a_{2xxx} - 2a_{2x}a_{2}^{2}a_{1xx} = 0
\end{eqnarray}

Since we require that the $a_{i}$ be real we decouple this final equation into the conditions

\begin{eqnarray}
&& a_{2t}a_{2x} - a_{2xt}a_{2} = 0 \\
&& a_{2}^{3}a_{1xxx} - 3a_{2}^{2}a_{2xx}a_{1x} - 4a_{2x}^{3}a_{1} + 5a_{1}a_{2}a_{2x}a_{2xx} + 4a_{2x}^{2}a_{2}a_{1x} \nonumber \\
&& - a_{2}^{2}a_{1}a_{2xxx} - 2a_{2x}a_{2}^{2}a_{1xx} = 0
\end{eqnarray}

With the aid of MAPLE we find that the previous system is solvable with solution given by

\begin{eqnarray}
a_{1}(x,t) &=& F_{4}(t)F_{2}(x)(c_{1} + c_{2}x) - c_{1}F_{4}(t)F_{2}(x)\int{\frac{x \ dx}{F_{2}(x)}} + c_{1}xF_{4}(t)F_{2}(x)\int{\frac{dx}{F_{2}(x)}} \\
a_{2}(x,t) &=& F_{2}(x)F_{3}(t)
\end{eqnarray}

By taking $f_{3} = 0$ and thus $\mathbb{X}_{2} = 0$ we are in fact removing $\eqref{DNLSb}$ as a requirement for $\mathbb{F}$ and $\mathbb{G}$ to have zero-curvature. Since $\eqref{DNLSa}$ and $\eqref{DNLSb}$ are complex conjugates of each other $r$ and $q$ satisfying one equation implies they satisfy the other. This completes the extended Estabrook-Wahlquist analysis
of our DNLS system.

\section{Conclusions}

In this paper, we have developed an extension of the well-known Estabrook-Whalquist method to algorithmically derive
generalizations of several well-known integrable systems with spatiotemporally varying coefficients. As discussed throughout,
this generalized Estabrook-Wahlquist technique recovers and systematizes the results \cite{Khawaja}-cite{17} obtained by 'guessing' generalizations
of the structures of the Lax Pairs for the corresponding constant-coefficient integrable systems.

Future work will consider further integrability properties of the various Lax-integrable systems derived here.

\appendix

\newpage
\setcounter{equation}{0}
\renewcommand{\theequation}{A.\arabic{equation}}

\section{Appendix: Intermediate Results for Fifth-Order Equation}

The intermediate results mentioned at the appropriate places in Section $3$ are given here, with the derivation
and the use of each detailed there. These intermediate results are:

\begin{eqnarray} 
&& \mathbb{X}_{1,t} + \mathbb{X}_{2,t}u - \mathbb{X}_{2}(a_{3}u^{2}u_{x} + a_{4}uu_{x} + a_{7}u + a_{8}u_{x}) - (a_{6}\mathbb{X}_{6})_{x}u_{x}^{2} + a_{1}\mathbb{X}_{4}u_{x}^{2} \nonumber \\
&& - (a_{6}\mathbb{X}_{2})_{xxxx}u_{x} - \frac{1}{2}(a_{1}\mathbb{X}_{2})_{x}u_{x}^{2} + \frac{1}{2}([\mathbb{X}_{2},(a_{6}\mathbb{X}_{2})_{x}])_{x}u_{x}^{2} + \frac{1}{2}(a_{2}\mathbb{X}_{2})_{x}u_{x}^{2} \nonumber \\
&& + (a_{6}\mathbb{X}_{4})_{xxx}u_{x} + ([\mathbb{X}_{1},(a_{6}\mathbb{X}_{2})_{x}])_{xx}u_{x} + ([\mathbb{X}_{2},(a_{6}\mathbb{X}_{2})_{x}])_{xx}uu_{x} - (a_{6}\mathbb{X}_{5})_{xx}u_{x} \nonumber \\
&& + (a_{1}\mathbb{X}_{4})_{x}uu_{x} - \frac{1}{2}(a_{6}\mathbb{X}_{6})_{x}u_{x}^{2} - ([\mathbb{X}_{2},(a_{6}\mathbb{X}_{4})_{x}])_{x}uu_{x} - (a_{5}\mathbb{X}_{2})_{xx}u_{x} - (a_{1}\mathbb{X}_{2})_{xx}uu_{x} \nonumber \\
&& + (a_{5}\mathbb{X}_{4})_{x}u_{x} + ([\mathbb{X}_{1},(a_{6}\mathbb{X}_{2})_{xx}])_{x}u_{x} + ([\mathbb{X}_{2},(a_{6}\mathbb{X}_{2})_{xx}])_{x}uu_{x} - ([\mathbb{X}_{1},(a_{6}\mathbb{X}_{4})_{x}])_{x}u_{x} \nonumber \\
&& - ([\mathbb{X}_{1},[\mathbb{X}_{1},(a_{6}\mathbb{X}_{2})_{x}]])_{x}u_{x} - ([\mathbb{X}_{1},[\mathbb{X}_{2},(a_{6}\mathbb{X}_{2})_{x}]])_{x}uu_{x} - ([\mathbb{X}_{2},[\mathbb{X}_{1},(a_{6}\mathbb{X}_{2})_{x}]])_{x}uu_{x} \nonumber \\
&& - ([\mathbb{X}_{2},[\mathbb{X}_{2},(a_{6}\mathbb{X}_{2})_{x}]])_{x}u^{2}u_{x} + (a_{6}\mathbb{X}_{10})_{x}u^{2}u_{x} - \mathbb{K}^{3}_{x} + ([\mathbb{X}_{2},(a_{6}\mathbb{X}_{2})_{x}])_{x}u_{x}^{2} + a_{6}\mathbb{X}_{9}u_{x}^{2} \nonumber \\
&& - [\mathbb{X}_{2},(a_{6}\mathbb{X}_{4})_{x}]u_{x}^{2} + [\mathbb{X}_{2},(a_{6}\mathbb{X}_{2})_{xx}]u_{x}^{2} + a_{6}\mathbb{X}_{8}u_{x}^{2} + (a_{6}\mathbb{X}_{9})_{x}uu_{x} + (a_{6}\mathbb{X}_{8})_{x}uu_{x} \nonumber \\
&& - [\mathbb{X}_{1},[\mathbb{X}_{2},(a_{6}\mathbb{X}_{2})_{x}]]u_{x}^{2} - [\mathbb{X}_{2},[\mathbb{X}_{1},(a_{6}\mathbb{X}_{2})_{x}]]u_{x}^{2} - (a_{1}\mathbb{X}_{2})_{x}u_{x}^{2} - (a_{6}\mathbb{X}_{6})_{xx}uu_{x} \nonumber \\
&& - 2[\mathbb{X}_{2},[\mathbb{X}_{2},(a_{6}\mathbb{X}_{2})_{x}]]uu_{x}^{2} + 2a_{6}\mathbb{X}_{10}uu_{x}^{2} - \mathbb{K}^{3}_{u}u_{x} + [\mathbb{X}_{1},(a_{6}\mathbb{X}_{5})_{x}]u_{x} - \frac{1}{2}a_{2}\mathbb{X}_{4}u_{x}^{2} \nonumber \\
&& + [\mathbb{X}_{1},(a_{6}\mathbb{X}_{2})_{xxx}]u_{x} + \frac{1}{2}a_{1}\mathbb{X}_{4}u_{x}^{2} - \frac{1}{2}[\mathbb{X}_{1},[\mathbb{X}_{2},(a_{6}\mathbb{X}_{2})_{x}]]u_{x}^{2} + [\mathbb{X}_{1},(a_{6}\mathbb{X}_{6})_{x}]uu_{x} \nonumber \\
&& - [\mathbb{X}_{1},(a_{6}\mathbb{X}_{4})_{xx}]u_{x} - [\mathbb{X}_{1},([\mathbb{X}_{1},(a_{6}\mathbb{X}_{2})_{x}])_{x}]u_{x} - [\mathbb{X}_{1},([\mathbb{X}_{2},(a_{6}\mathbb{X}_{2})_{x}])_{x}]uu_{x} \nonumber \\
&& - a_{1}\mathbb{X}_{5}uu_{x} + \frac{1}{2}a_{6}\mathbb{X}_{9}u_{x}^{2} + [\mathbb{X}_{1},[\mathbb{X}_{2},(a_{6}\mathbb{X}_{4})_{x}]]uu_{x} + [\mathbb{X}_{1},(a_{5}\mathbb{X}_{2})_{x}]u_{x} + (a_{6}\mathbb{X}_{7})_{x}u_{x} \nonumber \\
&& - a_{5}\mathbb{X}_{5}u_{x} - [\mathbb{X}_{1},[\mathbb{X}_{1},(a_{6}\mathbb{X}_{2})_{xx}]]u_{x} - [\mathbb{X}_{1},[\mathbb{X}_{2},(a_{6}\mathbb{X}_{2})_{xx}]]uu_{x} + [\mathbb{X}_{1},[\mathbb{X}_{1},(a_{6}\mathbb{X}_{4})_{x}]]u_{x} \nonumber \\
&& + [\mathbb{X}_{1},[\mathbb{X}_{1},[\mathbb{X}_{1},(a_{6}\mathbb{X}_{2})_{x}]]]u_{x} + [\mathbb{X}_{1},[\mathbb{X}_{1},[\mathbb{X}_{2},(a_{6}\mathbb{X}_{2})_{x}]]]uu_{x} + [\mathbb{X}_{1},[\mathbb{X}_{2},[\mathbb{X}_{1},(a_{6}\mathbb{X}_{2})_{x}]]]uu_{x} \nonumber \\
&& - a_{6}[\mathbb{X}_{1},\mathbb{X}_{7}]u_{x} - a_{6}[\mathbb{X}_{1},\mathbb{X}_{9}]uu_{x} - a_{6}[\mathbb{X}_{1},\mathbb{X}_{8}]uu_{x} + [\mathbb{X}_{2},[\mathbb{X}_{1},(a_{6}\mathbb{X}_{4})_{x}]]uu_{x} \nonumber \\
&& + [\mathbb{X}_{1},[\mathbb{X}_{2},[\mathbb{X}_{2},(a_{6}\mathbb{X}_{2})_{x}]]]u^{2}u_{x} - a_{6}[\mathbb{X}_{1},\mathbb{X}_{10}]u^{2}u_{x} + [\mathbb{X}_{1},\mathbb{K}^{3}] + [\mathbb{X}_{1},(a_{1}\mathbb{X}_{2})_{x}]uu_{x} \nonumber \\
&& + [\mathbb{X}_{2},(a_{6}\mathbb{X}_{2})_{xxx}]uu_{x} - \frac{1}{2}[\mathbb{X}_{2},[\mathbb{X}_{2},(a_{6}\mathbb{X}_{2})_{x}]]uu_{x}^{2} + [\mathbb{X}_{2},(a_{6}\mathbb{X}_{6})_{x}]u^{2}u_{x} + [\mathbb{X}_{2},(a_{6}\mathbb{X}_{5})_{x}]uu_{x} \nonumber \\
&& - [\mathbb{X}_{2},(a_{6}\mathbb{X}_{4})_{xx}]uu_{x} - [\mathbb{X}_{2},([\mathbb{X}_{1},(a_{6}\mathbb{X}_{2})_{x}])_{x}]uu_{x} - [\mathbb{X}_{2},([\mathbb{X}_{2},(a_{6}\mathbb{X}_{2})_{x}])_{x}]u^{2}u_{x} \nonumber \\
&& - a_{1}\mathbb{X}_{6}u^{2}u_{x} + \frac{1}{2}a_{6}\mathbb{X}_{10}uu_{x}^{2} + [\mathbb{X}_{2},[\mathbb{X}_{2},(a_{6}\mathbb{X}_{4})_{x}]]u^{2}u_{x} + [\mathbb{X}_{2},(a_{5}\mathbb{X}_{2})_{x}]uu_{x} \nonumber \\
&& - a_{5}\mathbb{X}_{6}uu_{x} - [\mathbb{X}_{2},[\mathbb{X}_{1},(a_{6}\mathbb{X}_{2})_{xx}]]uu_{x} - [\mathbb{X}_{2},[\mathbb{X}_{2},(a_{6}\mathbb{X}_{2})_{xx}]]u^{2}u_{x} \nonumber \\
&& + [\mathbb{X}_{2},[\mathbb{X}_{1},[\mathbb{X}_{1},(a_{6}\mathbb{X}_{2})_{x}]]]uu_{x} + [\mathbb{X}_{2},[\mathbb{X}_{1},[\mathbb{X}_{2},(a_{6}\mathbb{X}_{2})_{x}]]]u^{2}u_{x} + [\mathbb{X}_{2},(a_{1}\mathbb{X}_{2})_{x}]u^{2}u_{x} \nonumber \\
&& + [\mathbb{X}_{2},[\mathbb{X}_{2},[\mathbb{X}_{1},(a_{6}\mathbb{X}_{2})_{x}]]]u^{2}u_{x} - a_{6}[\mathbb{X}_{2},\mathbb{X}_{7}]uu_{x} - a_{6}[\mathbb{X}_{2},\mathbb{X}_{9}]u^{2}u_{x} - a_{6}[\mathbb{X}_{2},\mathbb{X}_{8}]u^{2}u_{x} \nonumber \\
&& + [\mathbb{X}_{2},[\mathbb{X}_{2},[\mathbb{X}_{2},(a_{6}\mathbb{X}_{2})_{x}]]]u^{3}u_{x} - a_{6}[\mathbb{X}_{2},\mathbb{X}_{10}]u^{3}u_{x} + [\mathbb{X}_{2},\mathbb{K}^{3}]u = 0, \label{KDV9}
\end{eqnarray}

\begin{eqnarray}
&& - \mathbb{X}_{2}(a_{3}u^{2} + a_{4}u + a_{8}) - (a_{6}\mathbb{X}_{2})_{xxxx} - \mathbb{K}^{3}_{u} + [\mathbb{X}_{1},(a_{6}\mathbb{X}_{5})_{x}] \nonumber \\
&& + (a_{6}\mathbb{X}_{4})_{xxx} + ([\mathbb{X}_{1},(a_{6}\mathbb{X}_{2})_{x}])_{xx} + ([\mathbb{X}_{2},(a_{6}\mathbb{X}_{2})_{x}])_{xx}u - (a_{6}\mathbb{X}_{5})_{xx} \nonumber \\
&& + (a_{1}\mathbb{X}_{4})_{x}u - ([\mathbb{X}_{2},(a_{6}\mathbb{X}_{4})_{x}])_{x}u - (a_{5}\mathbb{X}_{2})_{xx} - (a_{1}\mathbb{X}_{2})_{xx}u - a_{1}\mathbb{X}_{6}u^{2} \nonumber \\
&& + (a_{5}\mathbb{X}_{4})_{x} + ([\mathbb{X}_{1},(a_{6}\mathbb{X}_{2})_{xx}])_{x} + ([\mathbb{X}_{2},(a_{6}\mathbb{X}_{2})_{xx}])_{x}u - ([\mathbb{X}_{1},(a_{6}\mathbb{X}_{4})_{x}])_{x} \nonumber \\
&& - ([\mathbb{X}_{1},[\mathbb{X}_{1},(a_{6}\mathbb{X}_{2})_{x}]])_{x} - ([\mathbb{X}_{1},[\mathbb{X}_{2},(a_{6}\mathbb{X}_{2})_{x}]])_{x}u - ([\mathbb{X}_{2},[\mathbb{X}_{1},(a_{6}\mathbb{X}_{2})_{x}]])_{x}u \nonumber \\
&& + (a_{6}\mathbb{X}_{9})_{x}u + (a_{6}\mathbb{X}_{8})_{x}u - (a_{6}\mathbb{X}_{6})_{xx}u + [\mathbb{X}_{1},(a_{6}\mathbb{X}_{2})_{xxx}] + [\mathbb{X}_{1},(a_{6}\mathbb{X}_{6})_{x}]u \nonumber \\
&& - [\mathbb{X}_{1},(a_{6}\mathbb{X}_{4})_{xx}] - [\mathbb{X}_{1},([\mathbb{X}_{1},(a_{6}\mathbb{X}_{2})_{x}])_{x}] - [\mathbb{X}_{1},([\mathbb{X}_{2},(a_{6}\mathbb{X}_{2})_{x}])_{x}]u \nonumber \\
&& - a_{1}\mathbb{X}_{5}u + [\mathbb{X}_{1},[\mathbb{X}_{2},(a_{6}\mathbb{X}_{4})_{x}]]u + [\mathbb{X}_{1},(a_{5}\mathbb{X}_{2})_{x}] + (a_{6}\mathbb{X}_{7})_{x} + [\mathbb{X}_{2},[\mathbb{X}_{2},(a_{6}\mathbb{X}_{4})_{x}]]u^{2} \nonumber \\
&& - a_{5}\mathbb{X}_{5} - [\mathbb{X}_{1},[\mathbb{X}_{1},(a_{6}\mathbb{X}_{2})_{xx}]] - [\mathbb{X}_{1},[\mathbb{X}_{2},(a_{6}\mathbb{X}_{2})_{xx}]]u + [\mathbb{X}_{1},[\mathbb{X}_{1},(a_{6}\mathbb{X}_{4})_{x}]] \nonumber \\
&& + [\mathbb{X}_{1},[\mathbb{X}_{1},[\mathbb{X}_{1},(a_{6}\mathbb{X}_{2})_{x}]]] + [\mathbb{X}_{1},[\mathbb{X}_{1},[\mathbb{X}_{2},(a_{6}\mathbb{X}_{2})_{x}]]]u + [\mathbb{X}_{1},[\mathbb{X}_{2},[\mathbb{X}_{1},(a_{6}\mathbb{X}_{2})_{x}]]]u \nonumber \\
&& - a_{6}[\mathbb{X}_{1},\mathbb{X}_{7}] - a_{6}[\mathbb{X}_{1},\mathbb{X}_{9}]u - a_{6}[\mathbb{X}_{1},\mathbb{X}_{8}]u + [\mathbb{X}_{2},[\mathbb{X}_{1},(a_{6}\mathbb{X}_{4})_{x}]]u \nonumber \\
&& + [\mathbb{X}_{2},(a_{6}\mathbb{X}_{2})_{xxx}]u + [\mathbb{X}_{2},(a_{6}\mathbb{X}_{6})_{x}]u^{2} + [\mathbb{X}_{2},(a_{6}\mathbb{X}_{5})_{x}]u + [\mathbb{X}_{1},(a_{1}\mathbb{X}_{2})_{x}]u \nonumber \\
&& - [\mathbb{X}_{2},(a_{6}\mathbb{X}_{4})_{xx}]u - [\mathbb{X}_{2},([\mathbb{X}_{1},(a_{6}\mathbb{X}_{2})_{x}])_{x}]u - [\mathbb{X}_{2},([\mathbb{X}_{2},(a_{6}\mathbb{X}_{2})_{x}])_{x}]u^{2} \nonumber \\
&& - a_{5}\mathbb{X}_{6}u - [\mathbb{X}_{2},[\mathbb{X}_{1},(a_{6}\mathbb{X}_{2})_{xx}]]u - [\mathbb{X}_{2},[\mathbb{X}_{2},(a_{6}\mathbb{X}_{2})_{xx}]]u^{2} + [\mathbb{X}_{2},(a_{5}\mathbb{X}_{2})_{x}]u \nonumber \\
&& + [\mathbb{X}_{2},[\mathbb{X}_{1},[\mathbb{X}_{1},(a_{6}\mathbb{X}_{2})_{x}]]]u + [\mathbb{X}_{2},[\mathbb{X}_{1},[\mathbb{X}_{2},(a_{6}\mathbb{X}_{2})_{x}]]]u^{2} + [\mathbb{X}_{2},(a_{1}\mathbb{X}_{2})_{x}]u^{2} \nonumber \\
&& + [\mathbb{X}_{2},[\mathbb{X}_{2},[\mathbb{X}_{1},(a_{6}\mathbb{X}_{2})_{x}]]]u^{2} - a_{6}[\mathbb{X}_{2},\mathbb{X}_{7}]u - a_{6}[\mathbb{X}_{2},\mathbb{X}_{9}]u^{2} - a_{6}[\mathbb{X}_{2},\mathbb{X}_{8}]u^{2} = 0, \label{KDV13}
\end{eqnarray}

\noindent
and

\begin{eqnarray}
\mathbb{K}^{3} &=& - \frac{1}{3}a_{3}\mathbb{X}_{2}u^{3} - \frac{1}{2}a_{4}\mathbb{X}_{2}u^{2} - a_{8}\mathbb{X}_{2}u - (a_{6}\mathbb{X}_{2})_{xxxx}u + [\mathbb{X}_{1},(a_{6}\mathbb{X}_{5})_{x}]u - \frac{1}{2}a_{6}[\mathbb{X}_{2},\mathbb{X}_{7}]u^{2} \nonumber \\
&& + (a_{6}\mathbb{X}_{4})_{xxx}u + ([\mathbb{X}_{1},(a_{6}\mathbb{X}_{2})_{x}])_{xx}u + \frac{1}{2}([\mathbb{X}_{2},(a_{6}\mathbb{X}_{2})_{x}])_{xx}u^{2} - (a_{6}\mathbb{X}_{5})_{xx}u \nonumber \\
&& + \frac{1}{2}(a_{1}\mathbb{X}_{4})_{x}u^{2} - \frac{1}{2}([\mathbb{X}_{2},(a_{6}\mathbb{X}_{4})_{x}])_{x}u^{2} - (a_{5}\mathbb{X}_{2})_{xx}u - \frac{1}{2}(a_{1}\mathbb{X}_{2})_{xx}u^{2} - \frac{1}{3}a_{1}\mathbb{X}_{6}u^{3} \nonumber \\
&& + (a_{5}\mathbb{X}_{4})_{x}u + ([\mathbb{X}_{1},(a_{6}\mathbb{X}_{2})_{xx}])_{x}u + \frac{1}{2}([\mathbb{X}_{2},(a_{6}\mathbb{X}_{2})_{xx}])_{x}u^{2} - ([\mathbb{X}_{1},(a_{6}\mathbb{X}_{4})_{x}])_{x}u \nonumber \\
&& - ([\mathbb{X}_{1},[\mathbb{X}_{1},(a_{6}\mathbb{X}_{2})_{x}]])_{x}u - \frac{1}{2}([\mathbb{X}_{1},[\mathbb{X}_{2},(a_{6}\mathbb{X}_{2})_{x}]])_{x}u^{2} - \frac{1}{2}([\mathbb{X}_{2},[\mathbb{X}_{1},(a_{6}\mathbb{X}_{2})_{x}]])_{x}u^{2} \nonumber \\
&& + \frac{1}{2}(a_{6}\mathbb{X}_{9})_{x}u^{2} + \frac{1}{2}(a_{6}\mathbb{X}_{8})_{x}u^{2} - \frac{1}{2}(a_{6}\mathbb{X}_{6})_{xx}u^{2} + [\mathbb{X}_{1},(a_{6}\mathbb{X}_{2})_{xxx}]u + \frac{1}{2}[\mathbb{X}_{1},(a_{6}\mathbb{X}_{6})_{x}]u^{2} \nonumber \\
&& - [\mathbb{X}_{1},(a_{6}\mathbb{X}_{4})_{xx}]u - [\mathbb{X}_{1},([\mathbb{X}_{1},(a_{6}\mathbb{X}_{2})_{x}])_{x}]u - \frac{1}{2}[\mathbb{X}_{1},([\mathbb{X}_{2},(a_{6}\mathbb{X}_{2})_{x}])_{x}]u^{2} - \frac{1}{2}a_{1}\mathbb{X}_{5}u^{2} \nonumber \\
&& + \frac{1}{2}[\mathbb{X}_{1},[\mathbb{X}_{2},(a_{6}\mathbb{X}_{4})_{x}]]u^{2} + [\mathbb{X}_{1},(a_{5}\mathbb{X}_{2})_{x}]u + (a_{6}\mathbb{X}_{7})_{x}u + \frac{1}{3}[\mathbb{X}_{2},[\mathbb{X}_{2},(a_{6}\mathbb{X}_{4})_{x}]]u^{3} \nonumber \\
&& - a_{5}\mathbb{X}_{5}u - [\mathbb{X}_{1},[\mathbb{X}_{1},(a_{6}\mathbb{X}_{2})_{xx}]]u - \frac{1}{2}[\mathbb{X}_{1},[\mathbb{X}_{2},(a_{6}\mathbb{X}_{2})_{xx}]]u^{2} + [\mathbb{X}_{1},[\mathbb{X}_{1},(a_{6}\mathbb{X}_{4})_{x}]]u \nonumber \\
&& + [\mathbb{X}_{1},[\mathbb{X}_{1},[\mathbb{X}_{1},(a_{6}\mathbb{X}_{2})_{x}]]]u + \frac{1}{2}[\mathbb{X}_{1},[\mathbb{X}_{1},[\mathbb{X}_{2},(a_{6}\mathbb{X}_{2})_{x}]]]u^{2} + \frac{1}{2}[\mathbb{X}_{1},[\mathbb{X}_{2},[\mathbb{X}_{1},(a_{6}\mathbb{X}_{2})_{x}]]]u^{2} \nonumber \\
&& - a_{6}[\mathbb{X}_{1},\mathbb{X}_{7}]u - \frac{1}{2}a_{6}[\mathbb{X}_{1},\mathbb{X}_{9}]u^{2} - \frac{1}{2}a_{6}[\mathbb{X}_{1},\mathbb{X}_{8}]u^{2} + \frac{1}{2}[\mathbb{X}_{2},[\mathbb{X}_{1},(a_{6}\mathbb{X}_{4})_{x}]]u^{2} \nonumber \\
&& + \frac{1}{2}[\mathbb{X}_{2},(a_{6}\mathbb{X}_{2})_{xxx}]u^{2} + \frac{1}{3}[\mathbb{X}_{2},(a_{6}\mathbb{X}_{6})_{x}]u^{3} + \frac{1}{2}[\mathbb{X}_{2},(a_{6}\mathbb{X}_{5})_{x}]u^{2} + \frac{1}{2}[\mathbb{X}_{1},(a_{1}\mathbb{X}_{2})_{x}]u^{2} \nonumber \\
&& - \frac{1}{2}[\mathbb{X}_{2},(a_{6}\mathbb{X}_{4})_{xx}]u^{2} - \frac{1}{2}[\mathbb{X}_{2},([\mathbb{X}_{1},(a_{6}\mathbb{X}_{2})_{x}])_{x}]u^{2} - \frac{1}{3}[\mathbb{X}_{2},([\mathbb{X}_{2},(a_{6}\mathbb{X}_{2})_{x}])_{x}]u^{3} \nonumber \\
&& - \frac{1}{2}a_{5}\mathbb{X}_{6}u^{2} - \frac{1}{2}[\mathbb{X}_{2},[\mathbb{X}_{1},(a_{6}\mathbb{X}_{2})_{xx}]]u^{2} - \frac{1}{3}[\mathbb{X}_{2},[\mathbb{X}_{2},(a_{6}\mathbb{X}_{2})_{xx}]]u^{3} + \frac{1}{2}[\mathbb{X}_{2},(a_{5}\mathbb{X}_{2})_{x}]u^{2} \nonumber \\
&& + \frac{1}{2}[\mathbb{X}_{2},[\mathbb{X}_{1},[\mathbb{X}_{1},(a_{6}\mathbb{X}_{2})_{x}]]]u^{2} + \frac{1}{3}[\mathbb{X}_{2},[\mathbb{X}_{1},[\mathbb{X}_{2},(a_{6}\mathbb{X}_{2})_{x}]]]u^{3} + \frac{1}{3}[\mathbb{X}_{2},(a_{1}\mathbb{X}_{2})_{x}]u^{3} \nonumber \\
&& + \frac{1}{3}[\mathbb{X}_{2},[\mathbb{X}_{2},[\mathbb{X}_{1},(a_{6}\mathbb{X}_{2})_{x}]]]u^{3} - \frac{1}{3}a_{6}[\mathbb{X}_{2},\mathbb{X}_{9}]u^{3} - \frac{1}{3}a_{6}[\mathbb{X}_{2},\mathbb{X}_{8}]u^{3} + \mathbb{X}_{0}(x,t) \label{KDV14}
\end{eqnarray}

\end{document}